\documentclass[prb,aps,floatfix,amsmath,twocolumn]{revtex4}

\usepackage{amsmath}
\usepackage{bbm}
\usepackage{epsfig}
\usepackage{array}
\usepackage{verbatim}
\usepackage{hyperref}
\usepackage{amssymb}
\usepackage{multirow}
\usepackage{graphicx}

\usepackage{color}
\usepackage{ulem}

\usepackage{float}

\newcommand{\re}[1]{(\ref{#1})}

\newcommand{\beg}{\begin{equation}}

\newcommand{\en}{\end{equation}}

\newcommand {\dis}{\displaystyle}

\newcommand{\eps}{\varepsilon}
\newcommand{\lam}{\lambda}

\newcommand{\eref}[1]{Eq.~(\ref{#1})}
\newcommand{\esref}[1]{Eqs.~(\ref{#1})}

\newcommand{\lf}{\left\lfloor}
\newcommand{\rf}{\right\rfloor}

\renewcommand{\emph}{\textit}

\newcommand{\iden}{{\mathbbm{1}}}

\graphicspath{ {imagesreduced/} }
\usepackage{braket}

\begin{document}

\title{Integrable matrix theory: Level statistics}
\author{Jasen A. Scaramazza$^1$, B. Sriram Shastry$^2$, Emil A. Yuzbashyan$^1$}
\date{\today}							

\affiliation{$^1$Center for Materials Theory, Department of Physics and Astronomy, Rutgers University, Piscataway, NJ 08854, USA\\
$^2$Physics Department, University of California, Santa Cruz, CA 95064, USA}

\begin{abstract}
We study level statistics in ensembles of integrable $N\times N$ matrices linear in a real parameter $x$. The matrix $H(x)$ is considered integrable if it has a prescribed number $n>1$ of linearly independent commuting partners $H^i(x)$ (integrals of motion) $\left[H(x),H^i(x)\right] = 0$, $\left[H^i(x), H^j(x)\right]$ = 0, for all $x$. In a recent work, we developed a basis-independent construction of $H(x)$ for any $n$ from which we derived the probability density function, thereby determining how to choose a typical integrable matrix from the ensemble. Here, we find that typical integrable matrices have Poisson statistics in the $N\to\infty$ limit provided $n$ scales at least as $\log{N}$; otherwise, they exhibit level repulsion. Exceptions to the Poisson case occur at isolated coupling values $x=x_0$ or when correlations are introduced between typically independent matrix parameters. However, level statistics cross over to Poisson at $ \mathcal{O}(N^{-0.5})$ deviations from these exceptions, indicating that non-Poissonian statistics characterize only subsets of measure zero in the parameter space. Furthermore, we present strong numerical evidence that ensembles of integrable matrices are stationary and ergodic with respect to nearest neighbor level statistics.
 \end{abstract}

\maketitle

\section{Introduction}
\label{sec:defining}

It is generally believed that the energy levels of integrable systems\cite{note2} follow a Poisson distribution\cite{jpoilblanc,jrabson,jberry,jrelano,jstockmann,jellegaard,jputtner}. For example, the probability that a normalized spacing between adjacent levels lies between $s$ and $s+ds$ is expected to be $P(s)ds=e^{-s}ds$.  In contrast, chaotic systems exhibit Wigner-Dyson statistics, with level repulsion $P(s)\propto s^2$ or $s$ at small $s$. Moreover, level statistics are often used as a litmus test for quantum integrability  even though there are integrable models that fail this test, e.g. the reduced BCS model\cite{jrelano} (which is a particular linear combination of commuting Gaudin Hamiltonians). In this work, we quantify when and why Poisson statistics occur in quantum integrable models, while also characterizing exceptional (non-Poisson)  behavior. 

Poisson statistics have been numerically verified on a case-by-case basis for some quantum integrable systems, including the Hubbard\cite{jpoilblanc} and Heisenberg\cite{jpoilblanc, jrabson} models. On the other hand, general or analytic results on the spectra of quantum integrable models are lacking, in part due to the absence of a  generally accepted unambiguous notion of quantum integrability,\cite{caux,yuzbashyan1} and in part because existing results usually apply to isolated models instead of members of statistical ensembles like random matrices\cite{mehta}. Notably, Berry and Tabor showed \cite{jberry} that level statistics in semiclassical integrable models are always Poissonian as long as the energy $E(n_1,n_2,\dots)$ is not a linear function of the quantum numbers $n_1,n_2,\dots$, i.e., the system cannot be represented as a collection of decoupled harmonic oscillators. As integrability is destroyed by perturbing the Hamiltonian, the statistics are expected to cross over from Poisson to Wigner-Dyson  at perturbation strengths as small as the inverse system size\cite{jrabson}.

\begin{figure}
\includegraphics[width=\linewidth]{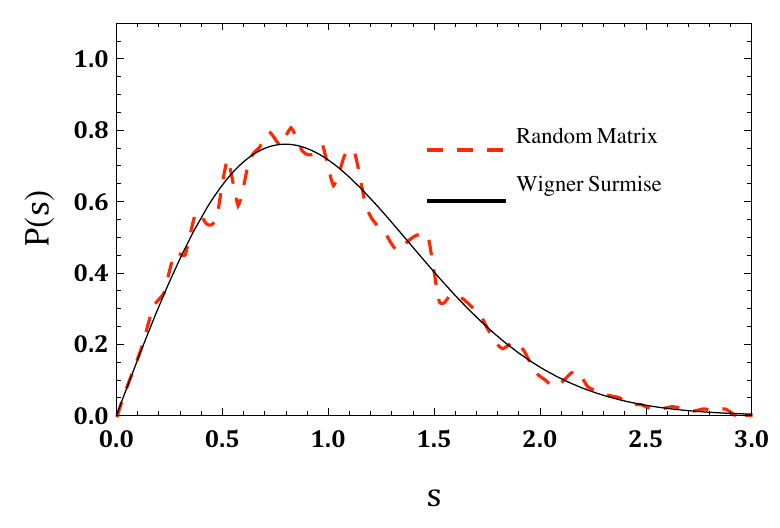}
\caption{(color online) The level spacing distribution of a $4000\times 4000$ random real symmetric matrix with entries chosen as independent random numbers from a normal distribution of mean 0 and off-diagonal variance 1/2 (diagonal variance of 1). Such a matrix belongs to the Gaussian orthogonal ensemble (GOE) of real symmetric matrices, studied in random matrix theory (RMT). The main feature of the spacing distribution here is its vanishing for small spacings, also known as level repulsion. The smooth curve is the Wigner surmise $P(s) = \frac{\pi}{2} s e^{-\frac{\pi}{4} s^2}$. See the integrable matrix case in Fig.~\ref{fig:Poisson}.}
\label{fig:WigSurmise}
\end{figure}

 Random matrix theory (RMT)\cite{mehta,jGuhr} captures level repulsion and other universal features of eigenvalue statistics in generic (non-integrable) Hamiltonians (see, e.g., Fig.~\ref{fig:WigSurmise}).  We recently proposed an integrable matrix theory\cite{ScYu} (IMT) to describe  eigenvalue statistics of integrable models. This theory  is based on a rigorous notion of quantum integrability  and provides ensembles of 
 integrable matrix Hamiltonians with any given number of integrals of motion (see below).  It is similar to RMT in that both are   ensemble theories equipped with rotationally invariant probability density functions. An important difference is that random matrices   do not represent realistic many-body models, while integrable ones correspond to actual integrable Hamiltonians. We therefore have  access not only to typical features, but also to  exceptional cases and are in a position to make definitive statements about the statistics   of quantum integrable models.
 Here, we study the nearest-neighbor level spacing distributions of the IMT ensembles.
 
The approach of Refs.~\onlinecite{yuzbashyan1,ScYu,owusu,owusu1,yuzbashyan,shastry,shastry1} to quantum integrability operates with $N\times N$ Hermitian matrices linear in a real parameter $x$. A matrix $H(x)=xT+V$ is called \textit{integrable}\cite{owusu,owusu1,note3} if it has a commuting partner $\widetilde{H}(x)=x\widetilde{T}+\widetilde{V}$ other than a linear combination of itself and the identity matrix and if $H(x)$ and  $\widetilde{H}(x)$ have no common $x$-independent symmetry, i.e., no $\Omega\ne c\iden$ such that $[\Omega, H(x)]=[\Omega,  \widetilde{H}(x)]=0$. Fixing the parameter-dependence makes the existence of  commuting partners a nontrivial condition, so that only a subset of measure zero among all Hermitian matrices of the form $xT+V$ are integrable\cite{owusu1}.

Further, integrable matrices fall into different classes (types) according to the number of independent integrals of motion. We say that $H(x)$ is a type-$M$  integrable matrix if there are precisely $n=N-M>1$ linearly independent $N\times N$ Hermitian matrices\cite{note1} $H^i(x)=xT^i+V^i$ with no common $x$-independent symmetry such that  
\beg
\left[H(x),H^i(x)\right] =0,\quad [H^i(x),H^j(x)] = 0,  
\label{comm}
\en
for all $x$ and $i,j=1,\dots, n$. 
A type-$M$ family of integrable matrices (\textit{integrable family}) is   an $n$-dimensional vector space\cite{note1}, where $H^i(x)$  provide a basis. The general member of the family is 
\beg
H(x)=\sum_{i=1}^n d_i H^i(x),
\label{introbasismat}
\en
 where $d_i$ are real numbers. The maximum possible value of $n$ is $n=N-1$, corresponding to type-1 or maximally commuting Hamiltonians.  
 
 Examples of well-known many-body Hamiltonians that fit into this definition of integrability are the Gaudin, 1D Hubbard and XXZ models, where $x$ corresponds to the external magnetic field, Hubbard $U$ and the anisotropy, respectively. Note, however, that these models have various $x$-independent symmetries, such as the $z$ component of the total spin, total momentum, etc. Taken at a given number of spins or sites, they break down into sectors (matrix blocks) characterized by  certain  parameter-independent  symmetry quantum numbers. Such blocks are integrable matrices according to our definition. For instance, the 1D Hubbard model on six sites with three spin up and three spin down electrons is a direct sum of integrable matrices of various types\cite{owusu1}. Sectors of Gaudin magnets, where the $z$-component of the total spin  differs by one from its maximum or minimum value (one spin flip), or, equivalently, the one Cooper pair sector of the  BCS model are type-1 \cite{owusu}, while other sectors are integrable matrices of higher types. 
 
 Prior work\cite{owusu,owusu1,shastry,shastry1,yuzbashyan1} constructed all type-1, 2,  3 integrable matrices and a certain subclass of arbitrary type-$M$, determined exact eigenvalues and eigenfunctions of these matrices, investigated   the number of level crossings as a function of size and type, and showed that type-1 integrable families satisfy the Yang-Baxter equation. This work is a continuation of Ref.~\onlinecite{ScYu} where we formulated a rotationally invariant parametrization of integrable matrices  and  derived  an appropriate probability density function (PDF) for the parameters, i.e., for ensembles of integrable matrices of any given type. The derivation  is similar to that in the RMT and is based on either maximizing the entropy of the PDF or, equivalently, postulating statistical independence of independent parameters  and rotational invariance of the PDF. Here, we use the results of Ref.~\onlinecite{ScYu} to generate and study numerically and analytically level spacing distributions in ensembles of integrable matrices of various types as well as in individual  matrices.


\begin{figure}
\includegraphics[width=\linewidth]{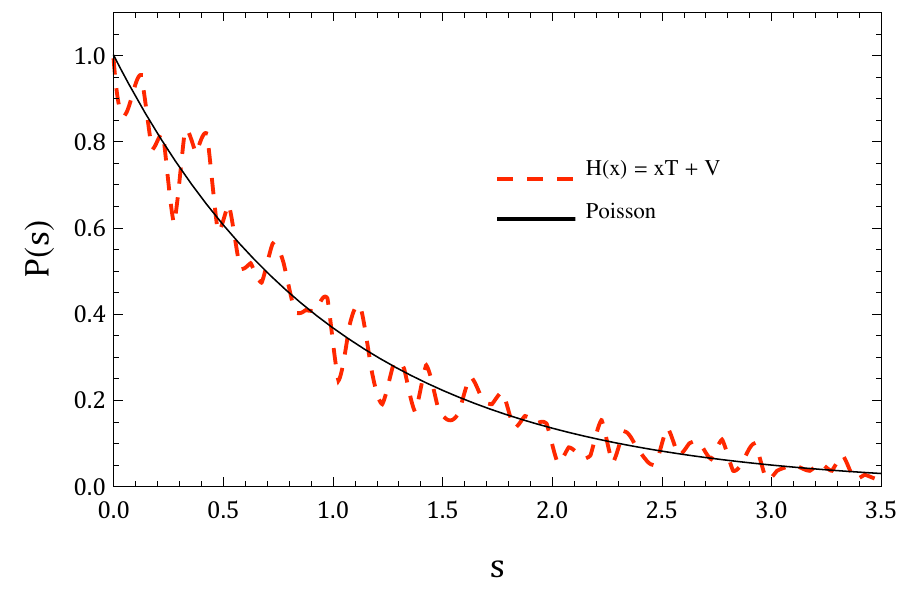}
\caption{(color online) The level spacing distribution for a $4000\times 4000$ real symmetric integrable matrix  $H(x) = xT + V$ at $x=1$. This particular matrix is a sum of 200 linearly independent matrices that commute for all values of the real parameter $x$. Note that the spacing distribution is maximized at $s = 0$, a feature known as level clustering. The smooth curve is a Poisson distribution, which is theorized to be typical of integrable matrices. Compare to the generic real symmetric matrix case in Fig.~\ref{fig:WigSurmise}.}
\label{fig:Poisson}
\end{figure}

Our main results are as follows.  For a generic choice of parameters, the level statistics of integrable matrices $H(x)$ are Poissonian in the limit of the Hilbert space size $N \to \infty$ if the number of conservation laws $n$ scales at least as $\log{N}$, see Fig.~\ref{fig:Poisson} for an example. Exceptions to Poisson statistics fall into two categories. First, it is always possible to construct an integrable matrix that has \textit{any} desired level spacing distribution at a given isolated value, $x=x_0$, of the coupling (or external field) parameter. For a typical type-1 matrix  there is always a single value of $x$ where the statistics are Wigner-Dyson. The distribution quickly crosses back over, however, to Poisson at deviations from $x_0$ of size $\delta x\sim N^{-0.5}$, with the crossover centered at $\delta x\sim N^{-1}$. Second, one obtains non-Poissonian distributions by introducing correlations among the ordinarily independent parameters characterizing an integrable matrix $H(x)$; the reduced BCS model falls into this category. The statistics again revert to Poisson at $\mathcal{O}(N^{-0.5})$ deviations from such correlations. We also show numerically that as $N \to \infty$, integrable matrix ensembles satisfy two distinct definitions of ergodicity with respect to the nearest-neighbor spacing distribution $P(s)$. Not only are the statistics of a single matrix representative of the entire ensemble, but the statistics of the $j$-th bulk spacing across the ensemble are independent of $j$.

In Sect.~\ref{Type1}, we present numerical results on the level statistics of type-1 matrices, defined to be integrable matrices $H(x)$ with the maximum number $n_{\mathrm{max}} = N - 1$ of linearly independent commuting partners. Section~\ref{sec:HigherTypesResults} contains numerical results for integrable matrices with $n \le n_{\mathrm{max}}$. We present our analytical justification of numerical results using perturbation theory in Sect.~\ref{exception}. Finally, we give numerical results on ergodicity in Sect.~\ref{sec:ergodicity}.

\section{Level statistics of type-1 integrable matrices}
\label{Type1}

\subsection{Type-1 families, primary parametrization}
\label{ss:Type1families}

Although our definition of integrable matrices encompasses the general Hermitian case, we restrict our focus in this work to real symmetric matrices. We begin with type-1 integrable $N \times N$ families which contain $N-1$ nontrivial commuting partners in addition to the scaled identity $  (c_1x + c_2)\iden$. Such matrices are the simplest to construct, for the parametrization of type-$M$ integrable families increases in complexity with $M$. Results on these higher types are deferred to Sect.~\ref{sec:HigherTypesResults}.

We first summarize the essential points of the basis-independent type-1 construction of Ref.~\onlinecite{ScYu} in order to arrive at the parametrization of \eref{eq:HType1Param} useful for numerical calculations. By considering linear combinations of the $N-1$ basis matrices, defined in \eref{introbasismat}, and the identity, one can prove that every type-1 family contains a particular integrable matrix $\Lambda(x)$ with rank-1 $T$-part
\beg
\Lambda(x) = x\ket{\gamma}\bra{\gamma}+E,  
\label{introreducedL}
\en
i.e., $[H(x),\Lambda(x)]=0$ for all $x$ and any $H(x)=xT+V$ in the family. There is an additional restriction  $[V,E]=0$, which follows from $\mathcal{O}(x^0)$ term in the commutator. It can be shown that the matrices $E$ and $V$ and the vector $\ket{\gamma}$ completely determine a given type-1 matrix $H(x)=xT+V$  modulo an additive constant proportional to the scaled identity.

If we consider any type-1 $H(x)$ in the shared eigenbasis of $E$ and $V$, we find that the matrix elements of $H(x)$ can be parametrized in terms of the $N$ eigenvalues $\eps_i$ of $E$, the $N$ eigenvalues $d_i$ of $V$, and the $N$ vector components $\gamma_i$  of $\ket{\gamma}$. Statistical arguments borrowed from RMT in Ref.~\onlinecite{ScYu} identify the $\eps_i$ and $d_i$ as two independent sets of eigenvalues drawn from the Gaussian orthogonal ensemble. The $\gamma_i$ are drawn from a $\delta(1-|\gamma|^2)$ distribution. With these parameters, \textit{any} $N\times N$ type-1 integrable matrix $H(x) = xT + V$ can be constructed in the following way:
\begin{equation}
\label{eq:HType1Param}
\begin{array}{l}
\dis [H(x)]_{ij} = x\gamma_i\gamma_j\dfrac{d_i - d_j}{\varepsilon_i - \varepsilon_j},\quad i\ne j, \\
\\
\dis \left[H(x)\right]_{jj} = d_j - x\sum_{k\ne j}{\gamma_k^2\dfrac{d_j - d_k}{\varepsilon_j - \varepsilon_k}}.
\end{array}
\end{equation}
We call \eref{eq:HType1Param} the ``primary" parametrization, which is given specifically in the basis where $V$ is diagonal and can be transformed into any other basis by an orthogonal transformation. Note that the quantities $d_j$ act as coefficients of linear combination of \textit{basis matrices} $H^i(x)$ defined by setting   $d_j = \delta_{ij}$ in \eref{eq:HType1Param}. Explicitly,   nonzero matrix elements of $H^i(x)$ are
\beg
\label{hi}
\begin{array}{l}
\dis [H^i(x)]_{ij}= [H^i(x)]_{ji}=  x\dfrac{\gamma_i\gamma_j}{\varepsilon_i - \varepsilon_j},\quad j\ne i, \\
\dis [H^i(x)]_{jj} =  - x\dfrac{\gamma_i^2}{\varepsilon_i - \varepsilon_j},\quad j\ne i, \\
\dis [H^i(x)]_{ii} =  1- x\sum_{k\ne i}\dfrac{\gamma_k^2}{\varepsilon_i-\varepsilon_k}.\\
\end{array}
\end{equation}
and
 \begin{equation}
H(x) = \sum_{i=1}^N{d_i H^i(x)}.
\label{eq:Type1Basis}
\end{equation}
From Eq.~(\ref{eq:Type1Basis}) we see that the $\varepsilon_i$ and $\gamma_i$ uniquely identify a type-1 commuting family whereas the choice of $d_i$ produces a given member of the family.

To describe the spectrum of $H(x)$, we introduce an additional $N$ parameters $\lambda_j = \lambda_j(x)$ determined by the following equation\cite{owusu}:
\begin{equation}
\label{eq:selfConsist}
\dfrac{1}{x}= \sum_{k=1}^N{\dfrac{\gamma^2_k}{\lambda_j - \varepsilon_k}}.
\end{equation}
One can graphically verify that for any non-degenerate choice of $\eps_k$ there are $N$ real solutions $\lambda_j$ to Eq.(\ref{eq:selfConsist}) that interlace the $\varepsilon_k$. The $N$ eigenvectors $v(x)$ and eigenvalues $\eta(x)$ of $H(x)$ are labeled by $\lambda_j$ and take the form
\begin{equation}
\label{eq:EigVecH}
\begin{array}{l}
\dis [v_{\lambda_j}(x)]_k = \dfrac{\gamma_k}{\lambda_j - \varepsilon_k},\quad \eta_{\lambda_j}(x) = x\sum_{i=1}^N{\dfrac{d_i \gamma_i^2}{\lambda_j - \varepsilon_i}}.
\end{array}
\end{equation}
The components of the (unnormalized) eigenvectors $v_{\lambda_j}(x)$ are independent of the choice of $d_i$ in \eref{eq:Type1Basis}, and are thus common to any member of the family defined by $\eps_k$ and $\gamma_k$.

\subsection{Universality of Poisson statistics}

Equipped with parametrizations of integrable matrix ensembles based on the number of commuting partners in a family, we can quantitatively outline both the origin and the robustness of Poisson statistics in these ensembles. We first explore the latter with numerical tests of the statistics of integrable matrices in Sects.~\ref{ssec:coupling} - \ref{ssec:basisansatz}. For clarity of exposition, the numerical results of Sects.~\ref{ssec:coupling}, \ref{ssec:correlation1} and \ref{ssec:basismatrices1} are demonstrated strictly for type-1 matrices. In Sect.~\ref{sec:HigherTypesResults}, we show that the same results apply generally to a construction of higher type integrable matrix families that by definition contain fewer than the maximum number of conservation laws. We present analytical considerations of numerical results in Sect.~\ref{exception}.

We emphasize that regardless of the choice of parameters we find Poisson level statistics in the overwhelming majority of cases, even near isolated points in parameter space with non-Poissonian statistics. For example, the least biased choice for $d_i$ in Sect.~\ref{ss:Type1families} enforces GOE statistics at $x=0$ since $H(0)=V$; by effecting a shift $x \to x+ x_0$, the equivalent invariant statement is that each type-1 matrix has a parameter value $x_0$ such that $H(x_0)$ has Wigner-Dyson statistics. Another exception to Poisson statistics is when $d_i$ and $\eps_i$ are correlated so that $d_i=f(\eps_i)$, a smooth function at least over almost the entire range of $\eps_i$. Nonetheless, as soon as we deviate from $x_0$ or $f(\eps_i)$, the results of Sects.~\ref{ssec:coupling} and \ref{ssec:correlation1} show that statistics quickly revert to Poisson at deviations scaling as $\delta \sim N^{-0.5}$ in the limit $N\to\infty$.

Generally, we find that random linear superpositions of basis matrices within a given integrable family are crucial for obtaining Poisson level statistics. Basis matrices themselves, defined in Eq.~(\ref{eq:Type1Basis}) for the primary type-1 construction and in Eq.~(\ref{him}) for more general integrable matrices, show non-Poissonian statistics with strong level repulsion. Such repulsion washes away, however, for $H(x)$ that are random linear combinations of sufficiently many basis matrices. We see this behavior in Sect.~\ref{ssec:basismatrices1} for all type-1 matrices, i.e., independent of the number $m$ of basis matrices (conservation laws) in linear combination as long as $m > \mathcal{O}(\log N)$.

We fit all spacing distributions $P(s)$ to the Brody\ function\cite{Brody} $P(s,\omega)$, where $\omega$ is the Brody parameter
\begin{equation}
\label{eq:Brody}
P(s,\omega) = a(\omega) s^{\omega} e^{-b(\omega) s^{\omega + 1}}.
\end{equation}
The distribution in Eq.~(\ref{eq:Brody}) has unit mean and norm with appropriate choices of constants $a(\omega)$ and $b(\omega)$. It interpolates between a Poisson distribution $P(s) = e^{-s}$ at $\omega = 0$ and the Wigner surmise $P(s) = \frac{\pi}{2} s e^{-\frac{\pi}{4} s^2}$ at $\omega = 1$, and hence is a convenient fitting function. The Brody parameter $\omega$ can take all values $\omega > -1$, which means it also can detect enhanced level clustering or repulsion.

Note, however, that the Wigner surmise is not the exact nearest neighbor spacing distribution of GOE matrices. One may therefore expect our numerics to produce an $\omega\ne 1$  for GOE matrices. Fig.~\ref{fig:Type1crossoverx}, where $\omega \approx .956$, shows that this is indeed the case. The exact distribution $P(s)$ can be found in Ref.~\onlinecite{mehta} and was originally derived by Gaudin in terms of a Fredholm determinant\cite{Gaudin61}. 
%
Using Ref.~\onlinecite{Borne} and a few lines of Mathematica code, we find that the same fitting procedure used for numerically generated matrices produces $\omega\approx0.957$. Note that it is important to exclude $P(0)=0$ in  the fitting procedure for numerically generated finite-sized matrices.

\subsection{Crossover in coupling parameter $x$}
\label{ssec:coupling}

\begin{figure}
\includegraphics[width=\linewidth]{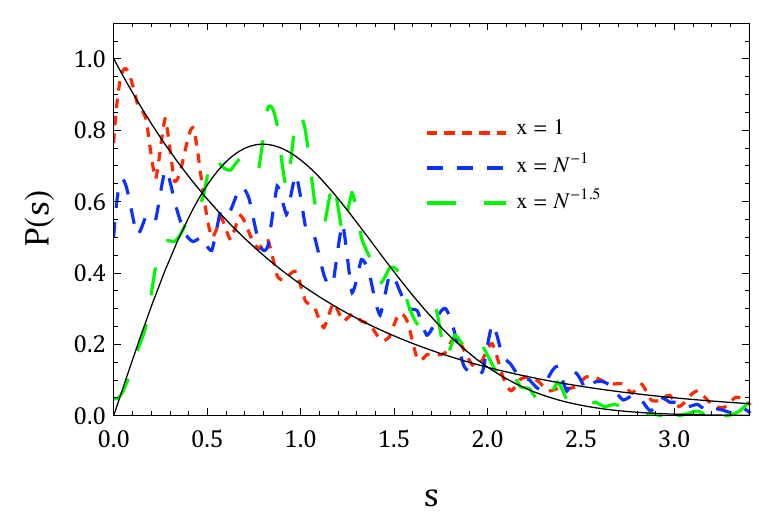}
\caption{(color online) Crossover in coupling $x$ of the level statistics of type-1 integrable $N\times N$ matrices $H(x) = xT + V$, $N = 4000$. See Sect.~\ref{ss:Type1families} for their parametrization. $V$ is a random matrix so that $H(x=0)$ has level repulsion. Each distribution contains the levels statistics of a single matrix $H(x)$ at a given value of $x$. Note that some level repulsion has set in by $x = N^{-1}$. Each numerical distribution is fit to the Brody function $P(s,\omega)$ from \eref{eq:Brody}; for couplings $x = \left(1, N^{-1}, N^{-1.5}\right)$ the fits give $\omega = \left(0.01, 0.30, 0.94\right)$, respectively. The solid lines are reference plots of a Poisson distribution $P(s) = e^{-s}$ and the Wigner Surmise $P(s) = \frac{\pi}{2}s e^{-\frac{\pi}{4}s^2}$. See Fig.~\ref{fig:Type1crossoverx} for more on this crossover.}
\label{fig:generalx}
\end{figure}

Here, we show that even if the statistics are   non-Poissonian at a given coupling value $x=x_0$ (we set $x_0 = 0$), level clustering is restored at small deviations from $x_0$. For any $N$, the matrices $T$ and $V$ each have eigenvalues that mostly lie on an $\mathcal{O}(1)$ interval centered about zero. We consider the primary type-1 construction encountered in Eq.~(\ref{eq:HType1Param}) and explore the level statistics of large matrices. In Fig.~\ref{fig:generalx}, we see qualitatively how the statistics change with $x$ when $N = 4000$. We find Poisson statistics at $x\sim 1$ until a crossover to level repulsion begins near $x=N^{-0.5}$ and ends near $x=N^{-1.5}$.

\begin{figure}
\includegraphics[width=\linewidth]{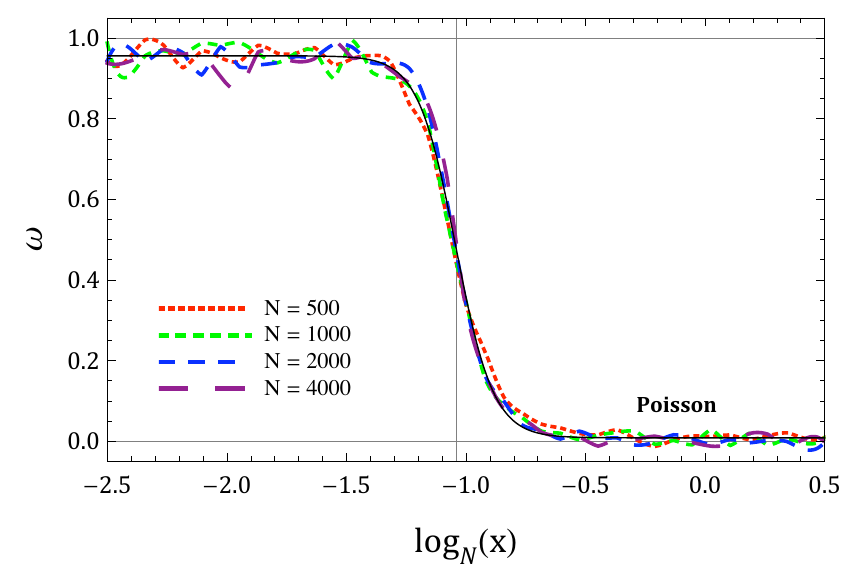}
\caption{(color online) Crossover in level statistics with variation of coupling parameter $x$ in type-1 integrable $N\times N$ matrices $H(x) = xT + V$, quantified by the Brody parameter $\omega(x,N)$ from \eref{eq:Brody}. The two important limits are $\omega = 0$ for Poisson statistics and $\omega = 1$ for random matrix (Wigner-Dyson) statistics. Each plotted value $\omega(x,N)$ is computed for the combined level spacing distribution of several matrices from the ensemble. We extract the crossover scale by fitting $\omega(x,N)$ to \eref{eq:BrodyFitFunction} (solid curve) to all curves simultaneously, where most notably $X_0 \sim -1$ for all $N$ investigated, indicating that crossovers to Poisson statistics are centered at that value for integrable matrices $H(x)$ when $H(x=0)$ has level repulsion. The middle of the crossover is indicated by a vertical line.}
\label{fig:Type1crossoverx}
\end{figure}

To verify that the crossover scaling inferred from Fig.~\ref{fig:generalx} is correct for all $N \gg 1$, in Fig.~\ref{fig:Type1crossoverx} we plot how the Brody parameter $\omega$ (see Eq.~(\ref{eq:Brody})) evolves with $x$ for various choices of $N$. It turns out that $\omega(x,N)$ can be fit to a relatively simple function, for any $N \gg 1$
\begin{equation}
\label{eq:BrodyFitFunction}
\omega(x,N) = \alpha - \beta\tanh\left({\dfrac{\log_N{x} - X_0}{Z}}\right).
\end{equation}
The numbers $(\alpha, \beta, X_0, Z)$ are fit parameters and take the values $(0.482,0.474,-1.04,0.157)$ in Fig.~\ref{fig:Type1crossoverx}. Most important is that for any $N \gg 1$ we find $X_0 \sim -1$, which solidifies our claim that the crossover occurs between $x \sim N^{-1.5}$ and $x \sim N^{-0.5}$. Analytical arguments explaining this scaling are given in Sect.~\ref{exception}.

\subsection{Correlations between matrix parameters}
\label{ssec:correlation1}

In the eigenbasis of $V$, our parametrization of integrable $N \times N$ matrices is given in terms of about $3N$ independent parameters (up to a change of basis). Through an explicit construction of the probability density function of integrable matrices obtained through basis-independent considerations, Ref.~\onlinecite{ScYu} shows that for a typical integrable matrix, $d_i$ and $\eps_i$ are indeed uncorrelated. We see in this section that if correlations are introduced between $\eps_i$ and $d_i$, the statistics become non-Poissonian. Small perturbations about these correlations, however, bring the statistics immediately back to Poisson. In this section, $x=1$ for all matrices considered.

\begin{figure}
\includegraphics[width=\linewidth]{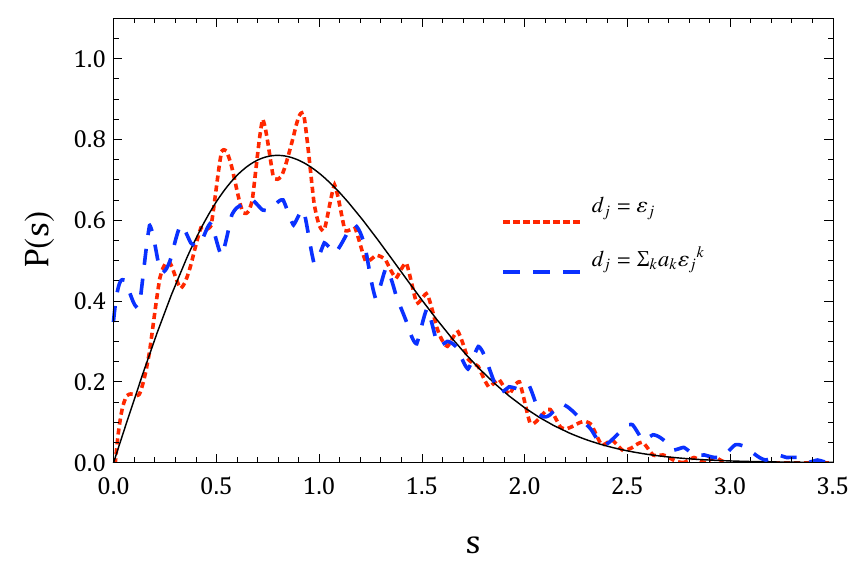}
\caption{(color online) Level statistics of two $N\times N$ type-1 integrable matrices $H(x) = xT + V$, $x=1$ and $N = 4000$, when correlations are introduced between $d_j$ and $\eps_j$ (see Eqs.~(\ref{eq:HType1Param}, \ref{eq:EigVecH}), and then Eq.~(\ref{BCS}) for an example). Note that in contrast to Fig.~\ref{fig:generalx}, these integrable matrices exhibit level repulsion even for $x = 1$. Each of the two curves is generated from a single matrix. One numerical curve corresponds to the case when $d_i = \eps_i$ and the other is when $d_i = \sum_{k=1}^4 A_k h_k(\eps_i)$, where $h_k (z)$ is the $k$-th order Hermite polynomial and $(A_1 , A_2, A_3, A_4 )= (2.3, 2.16, -1.46, 0.51)$, chosen randomly. Note that the polynomial dependence weakens the level repulsion as compared to the linear case. If higher order polynomials are included, the level repulsion eventually gives way to Poisson statistics. The solid curve is the Wigner surmise $P(s) = \frac{\pi}{2}s e^{-\frac{\pi}{4}s^2}$. See Fig.~\ref{fig:Type1CorrelationCrossover} for more on this behavior.}
 \label{fig:Type1CorrDE}
\end{figure}

Continuing with type-1 matrices in the primary parametrization, Eq.~(\ref{eq:HType1Param}), we recall that the eigenvalues $\eta_{\lambda_j}$ of such a matrix $H(x) = xT + V$ are given by \eref{eq:EigVecH},where the $\lambda_j = \lambda_j(x)$ are obtained from Eq.~(\ref{eq:selfConsist}). As we saw in  Sect.~\ref{ssec:coupling}, a typical choice of parameters will produce Poisson statistics, but this changes if we let $d_i$ be some smooth function of $\varepsilon_i$. The simplest case is shown in Fig.~\ref{fig:Type1CorrDE} for which $d_i = \varepsilon_i$. As discussed in Refs.~\onlinecite{ScYu,owusu1}, $H(x)$ for this choice of parameters  describes a sector of the reduced BCS model and, independently, a short range impurity in a weakly chaotic metallic quantum dot studied in Refs.~\onlinecite{aleiner,bogo}.

\begin{figure}
\includegraphics[width=\linewidth]{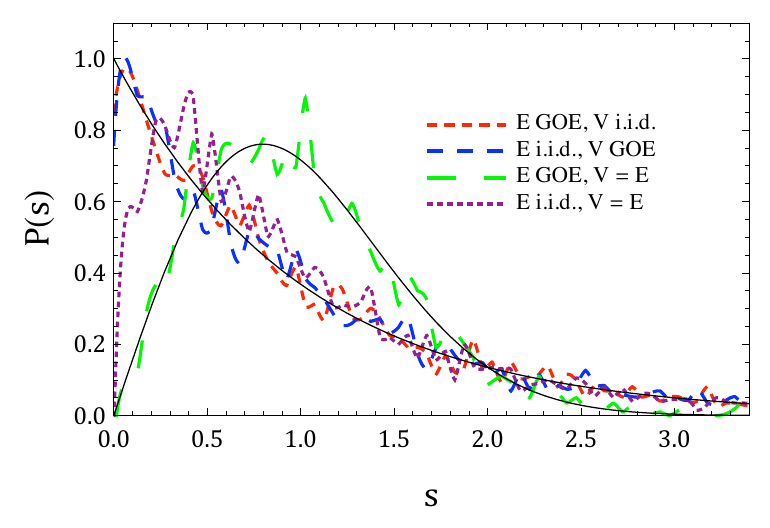}
\caption{(color online) Illustrating that conclusions drawn about correlations between $d_i$ (eigenvalues of $V$) and $\eps_i$ (eigenvalues of $E$) are independent of the particular choice of $\eps_i$. Pictured are four numerically generated nearest-neighbor spacing distributions $P(s)$ for $5000\times 5000$ type-1 matrices, $x=1$, when the $d_i$ and $\eps_i$ are either from a random matrix (GOE) or are independently and identically distributed numbers (i.i.d.) from a normal distribution. Each curve represents the level statistics of a single matrix chosen from the type-1 ensemble. Level repulsion survives in the two cases where $d_i$ and $\eps_i$ are correlated ($V=E$), even though the overall shape of $P(s)$ depends on whether $E$'s eigenvalues are GOE or i.i.d. numbers. The solid curves are the usual Poisson distribution $P(s) = e^{-s}$ and the Wigner surmise $P(s) = \frac{\pi}{2}se^{-\frac{\pi}{4}s^2}$. We do not include plots for different choices of $\gamma_i$, which do not affect the general character of the results.}

\label{diffparamchoices}
\end{figure}

The level repulsion for this case can be understood by a simple manipulation of Eq.~(\ref{eq:EigVecH}) when $d_i = \varepsilon_i$:
\begin{equation}
\begin{split}
\eta_{\lambda_j} &= x\sum_{i=1}^N{\dfrac{\varepsilon_i \gamma_i^2 + \lambda_j \gamma_i^2 - \lambda_j \gamma_i^2}{\lambda_j - \varepsilon_i}} \\
&= -x\sum_{i=1}^N{\gamma_i^2} + \lambda_j,
\end{split}
\label{BCS}
\end{equation}
where we used Eq.~(\ref{eq:selfConsist}). Then when $d_i = \eps_i$ the eigenvalues of $H(x)$ are just the $\lambda_j$ up to an additive constant. For the case when $\eps_i$ are random matrix eigenvalues,  Ref.~\onlinecite{aleiner} derives the joint probability density of the set $\{\eps_i,\lambda_j\}$ and Ref.~\onlinecite{bogo} demonstrates that the $\lambda_j$ are subject to the same level repulsion as the $\eps_i$. Note also that \eref{eq:selfConsist} implies  $\lam_j$ lie between consecutive $\eps_i$ and therefore the eigenvalues in \eref{BCS} can have no crossings  at any finite $x$. Numerically, we have found that $\lambda_j$ exhibit level repulsion for \textit{any} choice of $\eps_i$ (see Fig.~\ref{diffparamchoices}). Fig.~\ref{fig:Type1CorrDE} also shows the level repulsion induced when $d_i = \sum_{k=1}^4 A_k h_k(\eps_i)$, where $h_k(\eps_i)$ is the $k$-th order Hermite polynomial and $A_k$ are independent random numbers drawn from a normal distribution. In this case, the level repulsion is mitigated relative to the case of linear correlation. Sums to higher orders of $h_k(\eps_i)$ (or any higher order polynomial) will eventually bring the statistics back to Poisson.

\begin{figure}
\includegraphics[width=\linewidth]{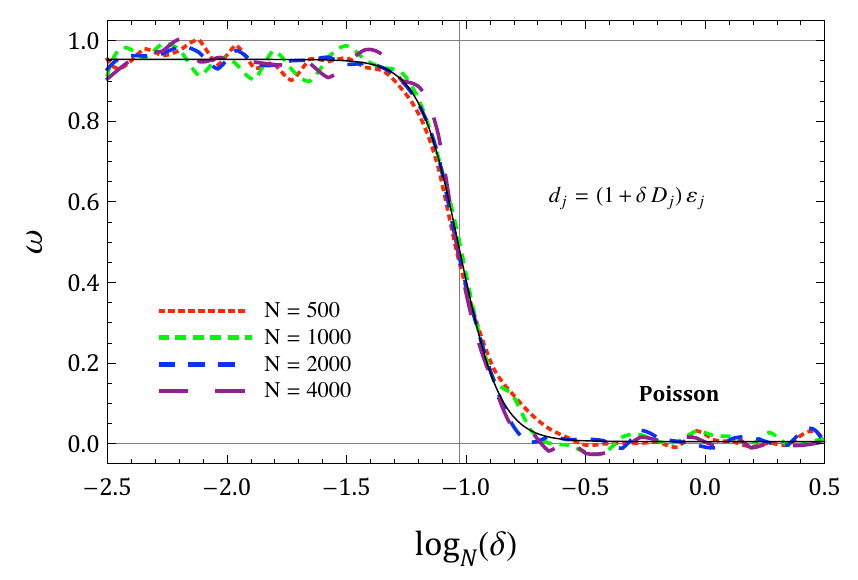}
\caption{(color online) Variation in the Brody parameter $\omega(\delta, N)$ when $d_i = \varepsilon_i(1 + \delta D_i)$ in the level statistics of $N\times N$ type-1 integrable matrices $H(x)$ for various $N$, $x = 1$. The number $\delta$ is a parameter controlling the size of the perturbation from correlation, and $D_i$ is an $\mathcal{O}(1)$ random number from a normal distribution. Note that the crossover in $\delta$ is very similar to that in $x$ shown in Fig.~\ref{fig:Type1crossoverx}. The numerical curves are fit to the function $\omega(\delta,N)$ given in \eref{eq:BrodyFitFunctionDelta} (solid curve), with a crossover centered at $X_0 \sim -1$, indicating that crossovers to Poisson statistics are centered at that value. Each plotted value $\omega(\delta,N)$ is computed for the combined level spacing distribution of several matrices from the ensemble. A vertical line indicates the center of the crossover on the plot. For a similar plot for nonlinear functions $d_i(\eps_i)$ see Fig.~\ref{fig:Type1CorrelationCrossover2}.}
 \label{fig:Type1CorrelationCrossover}
\end{figure}

\begin{figure}
\includegraphics[width=\linewidth]{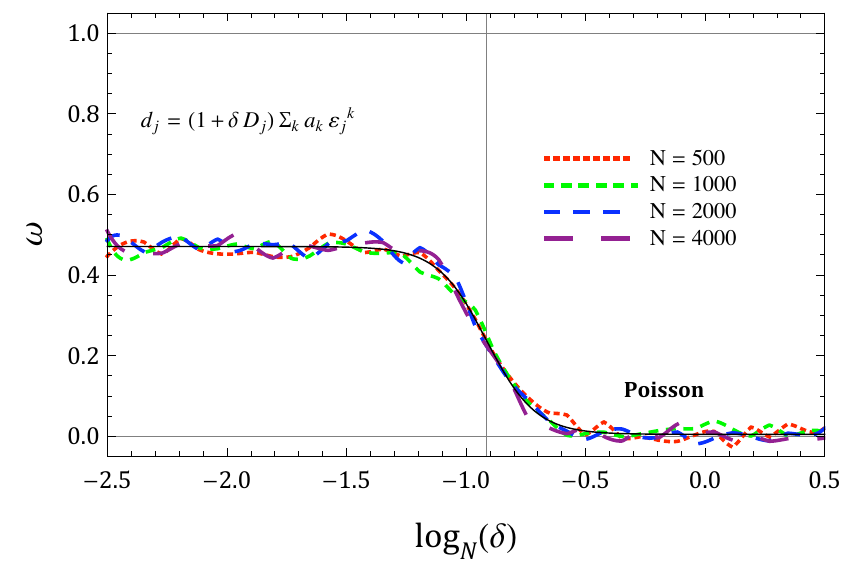}
\caption{(color online) Variation in the Brody parameter $\omega(\delta, N)$ when $d_i = \sum_{k=1}^4 A_k h_k(\eps_i)(1 + \delta D_i)$ in the level statistics of $N\times N$ type-1 integrable matrices $H(x=1)$ for various $N$.  Here $\delta$  quantifies the  deviation from the point $\delta=0$ where the parameters $d_i$ and $\varepsilon_i$ defining  the matrices are correlated, $D_i$  and $A_k$ are $\mathcal{O}(1)$ random numbers from a normal distribution, $h_k(z)$ is the $k$-th order Hermite polynomial. Each   $\omega(\delta,N)$ is computed for the combined level spacing distribution of several matrices from the ensemble. The crossover in $\delta$ is very similar to that in $x$  in Fig.~\ref{fig:Type1crossoverx} and   in $\delta$ for linear correlations in Fig.~\ref{fig:Type1CorrelationCrossover}.   Because the correlations are nonlinear, the level repulsion is   diminished in comparison to previous cases. Despite this, the crossover still demonstrates the same scaling -- fitting the  data to  $\omega(\delta,N)$ given in \eref{eq:BrodyFitFunctionDelta} (solid curve), with a crossover centered at $X_0 \sim -1$ (vertical line),  shows that  $\delta\propto N^{-0.5}$ is enough for statistics to revert to Poisson.}
\label{fig:Type1CorrelationCrossover2}
\end{figure}

We now investigate the stability of induced level repulsion in $H(x)$ when correlations between $d_i$ and $\eps_i$ are broken. In Fig.~\ref{fig:Type1CorrelationCrossover}, we let $d_i = \varepsilon_i (1 + \delta D_i)$ where $D_i$ is an $\mathcal{O}(1)$ random number from a normal distribution and $\delta$ is a number controlling the size of the perturbation. The crossover to Poisson statistics as $\delta$ increases is very similar to that in Fig.~\ref{fig:Type1crossoverx}, which shows the crossover with $x$. In fact, we can fit the Brody parameter $\omega(\delta,N)$ to 
\begin{equation}
\label{eq:BrodyFitFunctionDelta}
\omega(\delta,N) = \alpha - \beta\tanh\left({\dfrac{\log_N{\delta} - X_0}{Z}}\right).
\end{equation}
Note that Eq.~(\ref{eq:BrodyFitFunctionDelta}) is just Eq.~(\ref{eq:BrodyFitFunction}) with the substitution $x \to \delta$. We find that the crossover occurs over the range $N^{-1.5} \lesssim \delta \lesssim N^{-0.5}$, indicating that any perturbation to correlations will immediately destroy level repulsion as $N \to \infty$. In particular, Fig.~\ref{fig:Type1CorrelationCrossover} gives $(\alpha, \beta, X_0, Z) = (0.479,0.474,-1.03,0.169)$ for linear correlations. This scaling is not restricted to the case $d_i = \varepsilon_i$, as seen in Fig.~\ref{fig:Type1CorrelationCrossover2} where we again consider $d_i = \sum_{k=1}^4 A_k h_k(\eps_i)$ and find a similar crossover with $(\alpha, \beta, X_0, Z) = (0.237,0.233,-0.914,0.206)$.


\subsection{Basis matrices: how many conservation laws?}
\label{ssec:basismatrices1}

Here we demonstrate that in order to obtain Poisson statistics, the number $m$ of linearly independent conservation laws contained in an $N\times N$ integrable type-1 matrix can be much less than $N$. Consider a  combination of $m$ basis matrices $H^i(x)$ defined in Sect.~\ref{ss:Type1families}
\beg
H(x) = \sum_{i = 1}^{m}{d_i H^i(x)}, \quad m \le N-1.
\label{sumbasis}
\en
From the sum in \eref{sumbasis}, we can determine the number $m$ needed to obtain Poisson statistics. Individual basis matrices $H^i(x)$ will exhibit level repulsion, and it is only when an integrable matrix is formed from an \textit{uncorrelated} (w.r.t. $\eps_i$, see Sect.~\ref{ssec:correlation1}) linear combination of sufficiently many of them will we observe Poisson statistics. Level repulsion in this case can be qualitatively understood by reasoning that a basis matrix only ``contains" one nontrivial conservation law,  itself. More concretely, we see from  Eq.~(\ref{eq:EigVecH}) that the eigenvalues of   $H^i(x)$ are $x\gamma_i^2(\lambda_j-\varepsilon_i)^{-1}$, i.e., they are simple, mostly smooth functions of $\lambda_j$, which exhibit level repulsion.

\begin{figure}
\includegraphics[width=\linewidth]{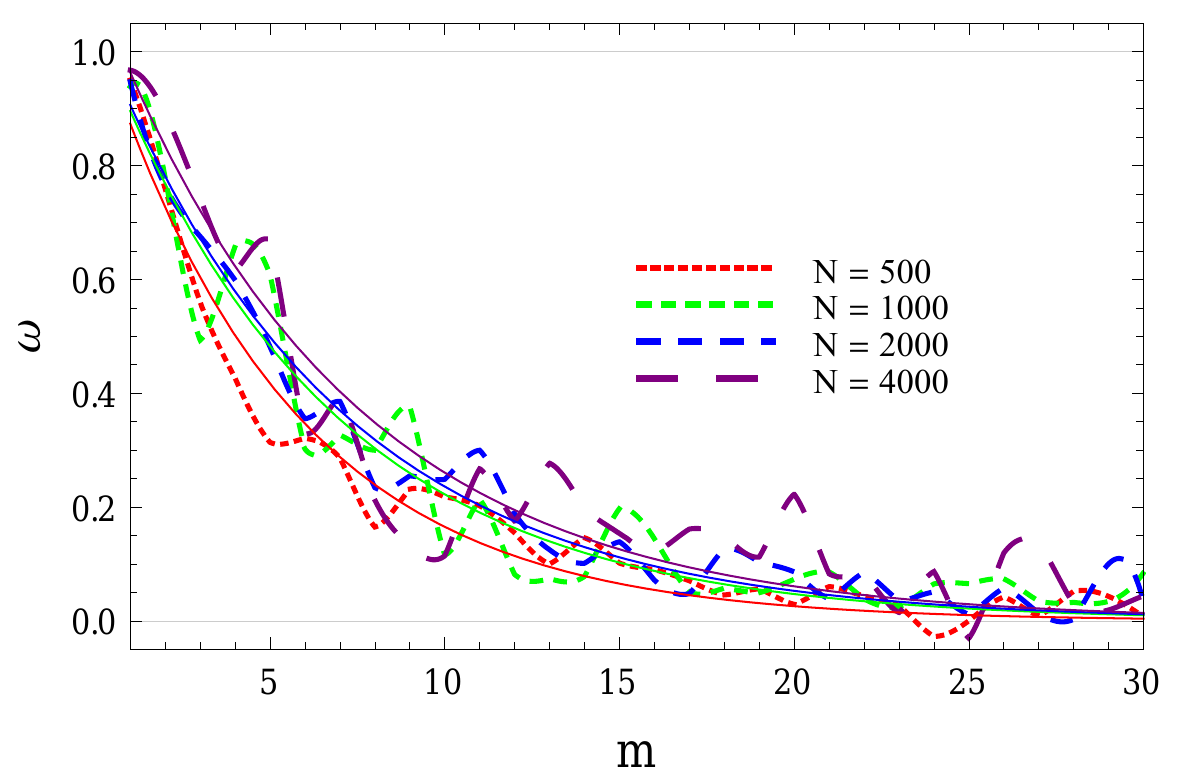}
\caption{(color online) The Brody parameter $\omega(m,N)$ (see \eref{eq:Brody}) vs. number $m$ of type-1 basis matrices $H^i(x)$  in linear combination $H(x) = \sum_{i=1}^m d_i H^i(x)$ for various $N$, $x=1$. The fits presume exponential decay and are expressed in terms of two parameters $(a,b)$ from Eq.~(\ref{eq:basisfit}). For $N = (500, 1000, 2000, 4000)$ we find the decay constant $b =(1.15,1.07,1.14,1.21)$, indicating that we only need $m_{\textrm{min}} \approx \log{N}$ conservation laws for Poisson statistics to emerge. Figs.~\ref{AnsatzBasis500} and \ref{AnsatzBasis2000} show similar plots for higher types. Each plotted $\omega(m,N)$ is computed for the combined level spacing distribution of several matrices from the ensemble.}
 \label{fig:Type1Basis}
\end{figure}

Fig. \ref{fig:Type1Basis} quantifies how many basis matrices $m$ (i.e., conservation laws) are needed for Poisson statistics as a function of $N$, the matrix size. We find numerically that the plots of the Brody parameter $\omega$ (see Eq.~(\ref{eq:Brody})) vs. the number $m$ of basis matrices in linear combination can be fit to a simple function
\begin{equation}
\omega(m,N) =  a \,\textrm{exp}\left[-\frac{b}{\log{N}}\,m\right],
\label{eq:basisfit}
\end{equation}
where $a$ and $b$ are real constants. The fact that for different values of $N$ we find that $b \sim 1$ supports the notion that we need only about $\log{N}$ conservation laws in order to induce Poisson statistics. We make this claim with caution because we only have data for $500 \le N \le 4000$, a range over which $\log{N}$ does not vary significantly. More precisely, Fig.~\ref{fig:Type1Basis} shows that having $m = \mathcal{O}(1)$ conservation laws is insufficient for inducing Poisson statistics, and that a useful upper bound on the lowest $m$ necessary for Poisson statistics is $m_{\textrm{min}} < \mathcal{O}(N^{\alpha})$ where $0 < \alpha < 0.20$. We obtain the factor of 0.20 by rewriting \eref{eq:basisfit} assuming the decay constant has power law dependence on $N$ instead of logarithmic dependence
\begin{equation}
\omega(m,N) =  a\, \textrm{exp}\left[-\frac{c}{N^{\alpha}}\,m\right].
\label{eq:basisfit2}
\end{equation}
Numerically we found that the parameter $b$ in \eref{eq:basisfit} satisifies $1.07\le b\le 1.21$ when $500\le N\le 4000$. By matching exponents between \eref{eq:basisfit2} and \eref{eq:basisfit} for $(b_1,N_1)=(1.21,500)$ and $(b_2,N_2)=(1.07,4000)$, we find a maximum exponent $\alpha = 0.198$.

\begin{figure}
\includegraphics[width=\linewidth]{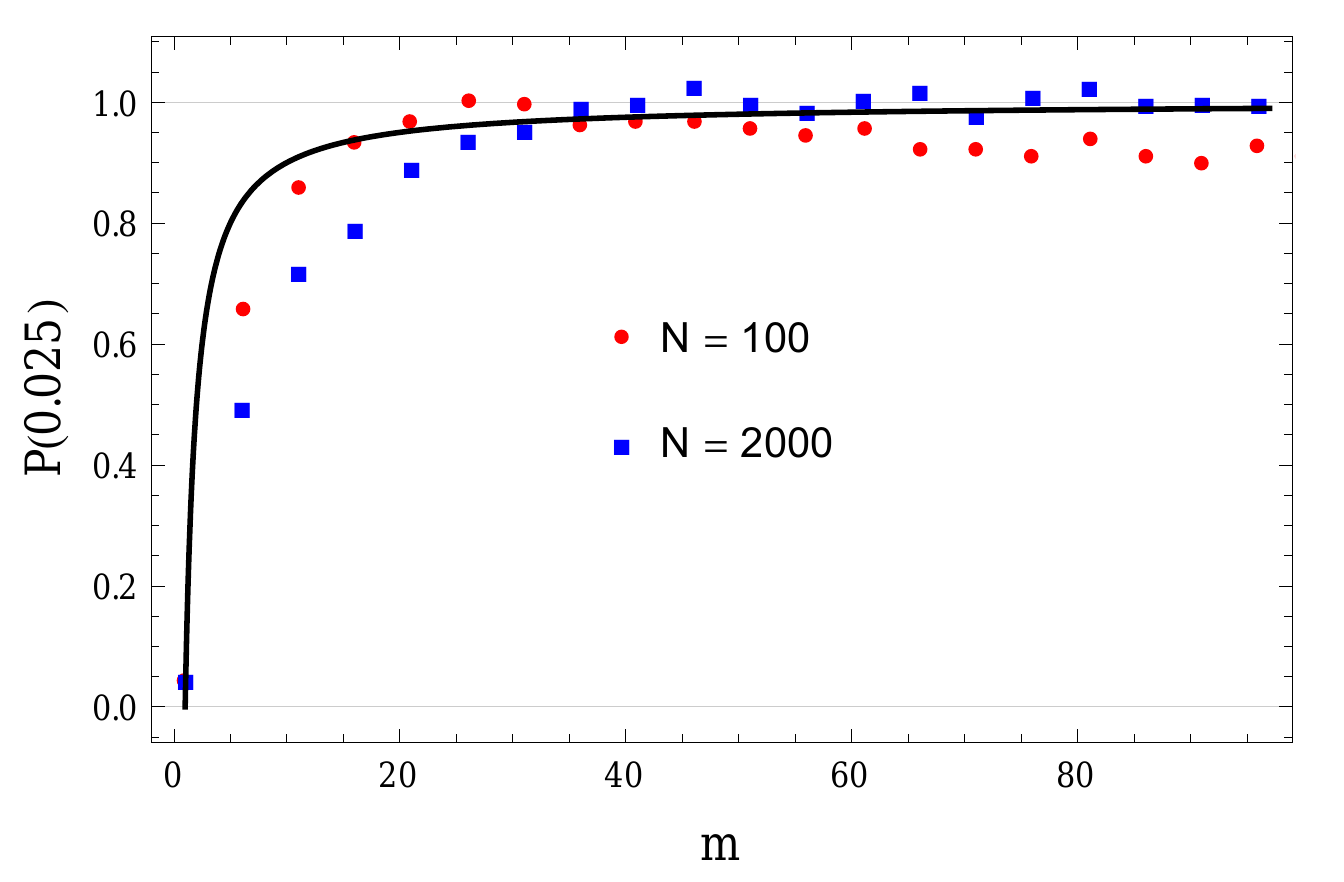}
\caption{(color online) Plot of numerically generated $P_m(0.025)$ for \textit{linear combinations} of $m$ type-1 basis matrices, \eref{sumbasis}, for $N=100$ and $N=2000$ at $x=1$. The solid curve gives the Rosenzweig-Porter prediction of $P_m(0)=1-1/m$ for \textit{superpositions} of $m$ independent random matrix spectra. Physically, the RP curve represents $P_m(0)$ for the combined spectra of $m$ blocks of different (parameter-independent) quantum numbers of a Hamiltonian. We note that although different mechanisms are involved in the RP and integrable matrix approach to Poisson statistics, the behavior of $P(0)$ is similar. This gives heuristic justification to why the existence of parameter-dependent conservation laws in $H(x)$ implies Poisson statistics. The sub-Poisson behavior for $N=100$ is a finite-size effect.}
 \label{RPcomp}
\end{figure}

The basis matrices $H^i(x)$ contained in any integrable $H(x)$ are linearly independent conservation laws. The observed dependence of $P(s)$ on the number $m$ of basis matrices in linear combination is reminiscent of the early work of Rosenzweig and Porter\cite{RP} (RP) on the nearest neighbor spacing distribution of superpositions of independent spectra. Although the spectra of basis matrices $H^i(x)$ are not strictly independent and are added together instead of superposed (``superposed'' here means ``combined into a single list''), we see the same qualitative behavior as described by RP: a single basis matrix has level repulsion, but a sufficiently large number combined have Poisson statistics. In the case of $m$ independent, superposed spectra with vanishing $P(0)$ that contribute equally to the mean level density, the value $P_m(0)$ of the superposed spectrum is given by the RP result
\beg
P_m(0) = 1-\frac{1}{m}.
\label{RPres}
\en
We see in Fig.~(\ref{RPcomp}) that $P_m(0)$ for $m$ basis matrices in linearly combination differs from the RP result for small $m$, as expected, but asymptotically approaches \eref{RPres} for large $m$ and large $N$. Thus it seems reasonable to conceptually understand the emergence of Poisson level statistics in integrable matrices $H(x)$ as arising from the existence of  conservation laws, whose spectra are  statistically independent for large $m$ and $N$.

Integrable matrix spectra are similar in structure to those of semiclassically integrable models studied by Berry and Tabor\cite{jberry}. Such spectra are also sums (or simple functions) of rigid spectra, and they have Poisson nearest-neighbor level statistics in the semiclassical limit.

Berry's work\cite{berry2} on semiclassical models shows that longer range spectral statistics of integrable and chaotic models deviate from the predictions of the Poisson ensemble\cite{poissonnote} and Gaussian random matrix theory, respectively. Similar behavior occurs in purely quantum systems\cite{seligman}. An example of such a long range statistic is $\Sigma^2(L)$, the spectral variance of the average number of eigenvalues contained in an interval of length $L$. For independent random numbers with unit mean spacing in an infinitely large spectrum, $\Sigma^2(L)=L$. For a given Hamiltonian, $\Sigma^2(L)$ will eventually saturate\cite{rigiditynote} at some $L_{\textrm{max}}$, which depends on the system's classical periodic orbits and the energy scale. 

\begin{figure}
\includegraphics[width=\linewidth]{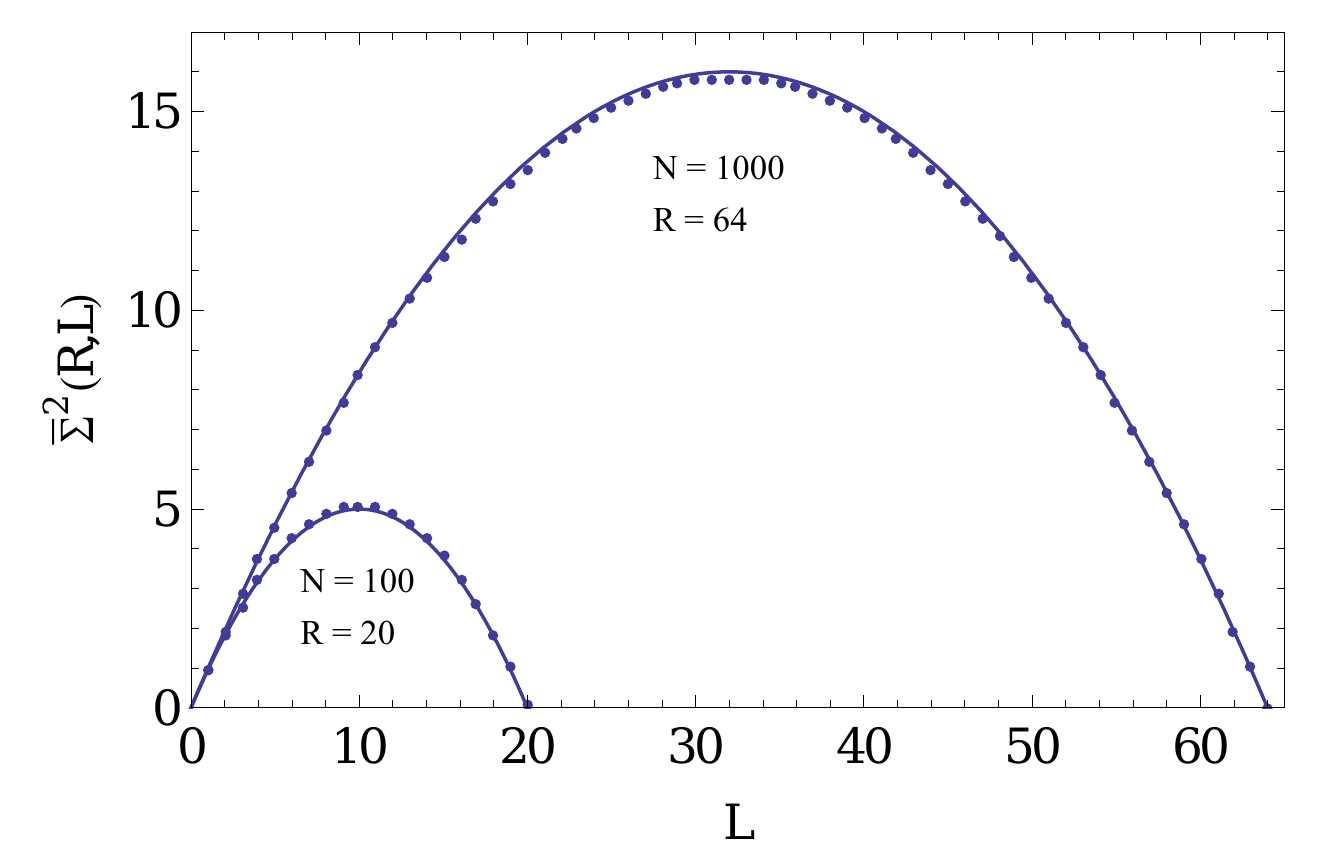}
\caption{(color online) Ensemble averaged number variance $\overline{\Sigma}^2(R,L)$ in $N\times N$ type-1 matrices $H(x) = xT+V$ at $x=1$ for $N=100$ and $N=1000$. In order to achieve a constant mean level spacing normalized to unity, we selected the middle $R=2\sqrt{N}$ eigenvalues from each matrix and used periodic boundary conditions on the list of eigenvalues. The results are in excellent agreement with the Poisson ensemble predictions (solid curves), given by \eref{rigidity}. There is no saturation on the ensemble average. We averaged over $10^4$ matrices for $N=100$ and $500$ matrices for $N=1000$.}
 \label{ensRig}
\end{figure}

We find no evidence of saturation of $\Sigma^2(L)$ in type-1 matrices on the ensemble average. Because we work with finite-size spectra, we compare numerically generated $\Sigma^2(L)$ to the corresponding Poisson ensemble averaged result for lists of $R$ independent numbers with unit mean spacing and periodic boundary conditions
\beg
\overline{\Sigma}^2(R,L) = L\left(1-\frac{L}{R}\right).
\label{rigidity}
\en
The overline indicates an average over the Poisson ensemble. Because numerical unfolding (see Appendix~\ref{appA}) introduces spurious effects in long range spectral observables, we instead average over small regions containing $R=2\sqrt{N}$ eigenvalues in the centers of $N\times N$ matrices where the level density is approximately constant. As seen in Fig.~\ref{ensRig}, the spectral variance of type-1 matrices satisfies \eref{rigidity}, even at relatively small $N$. 

\begin{figure}
\includegraphics[width=\linewidth]{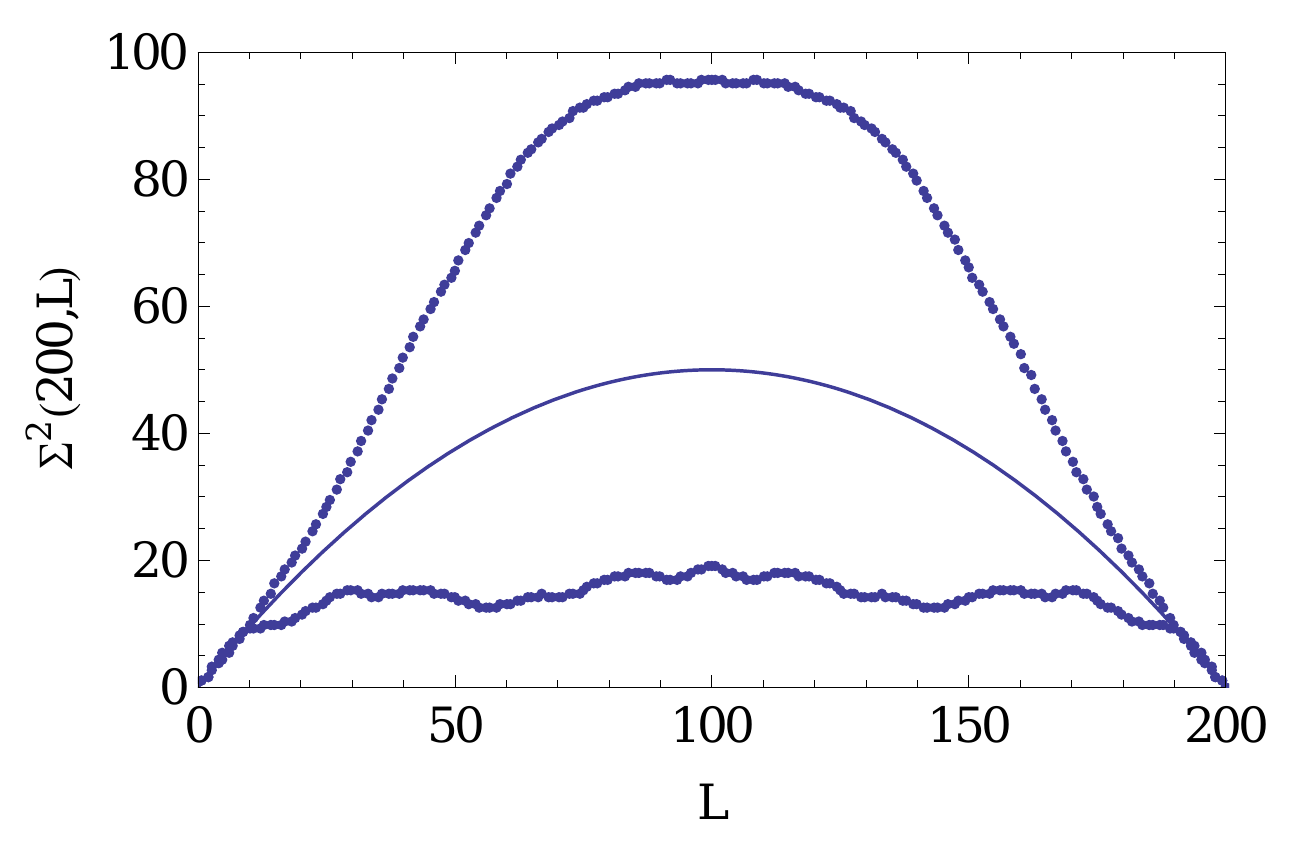}
\caption{(color online) Deviations from the Poisson ensemble average \eref{rigidity} (solid curve) of number variance $\overline{\Sigma}^2(200,L)$ from of two members of the Poisson ensemble. Shown are the number variances of two different lists of 200 independent numbers from a flat distribution in order to illustrate the large fluctuations of long-range spectral observables in the Poisson ensemble. See Fig.~\ref{specRigT1} for similar behavior in type-1 spectra.}
 \label{specRigPois}
\end{figure}

\begin{figure}
\includegraphics[width=\linewidth]{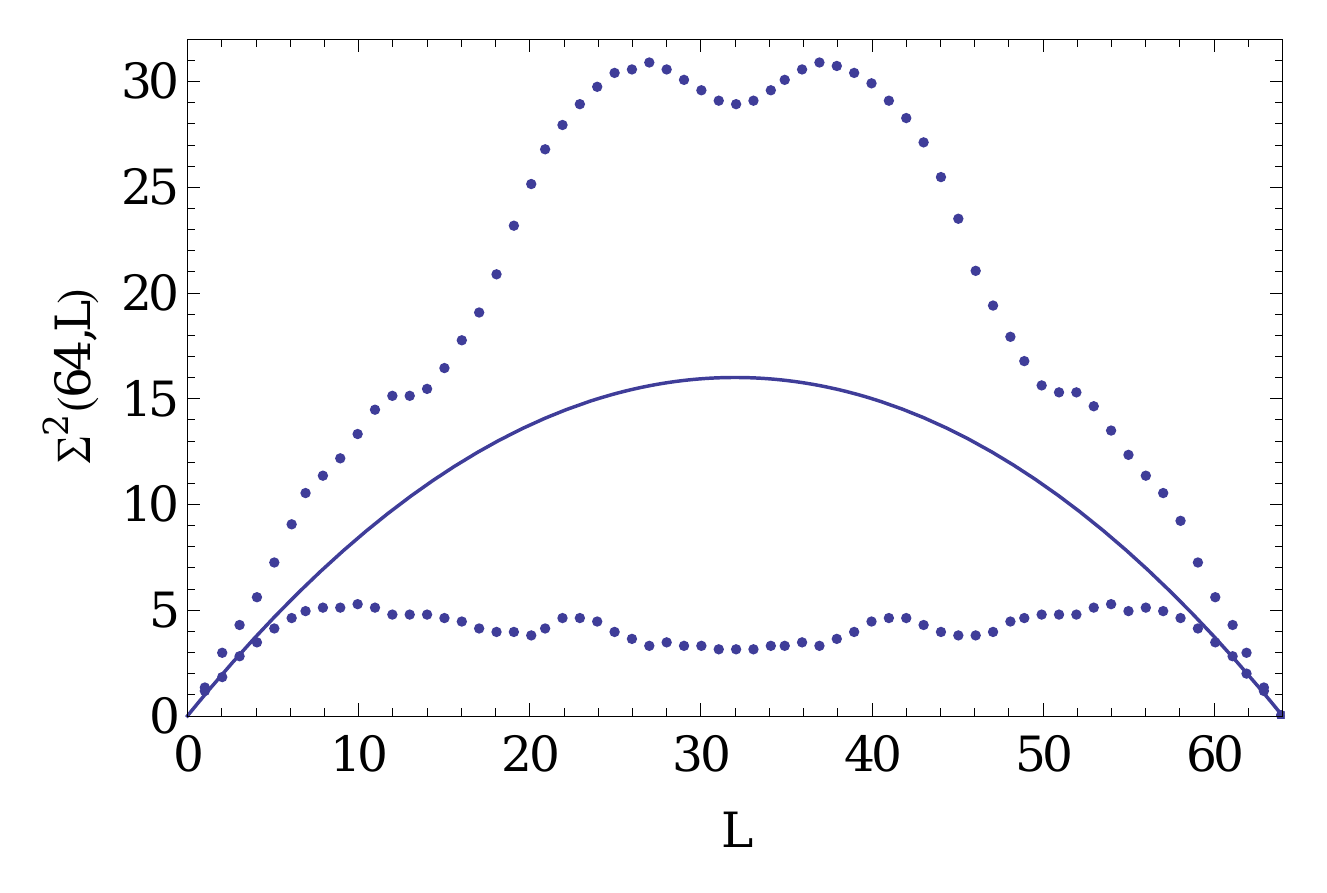}
\caption{(color online) Deviations from the Poisson ensemble average \eref{rigidity} (solid curve) of number variance $\overline{\Sigma}^2(64,L)$ from of two members of the $N=1000$ type-1 ensemble. Shown are the number variances of two matrices used in the ensemble average of Fig.~\ref{ensRig}. The saturation observed in the more rigid of the two spectra is reminiscent of that seen in members of the Poisson ensemble, see Fig.~\ref{specRigPois}.}
 \label{specRigT1}
\end{figure}

While there is no saturation on the ensemble average, $\Sigma^2(R,L)$ in the Poisson ensemble has large fluctuations for $L\sim R/2$. Figs.~\ref{specRigPois} and \ref{specRigT1} show how individual members of the Poisson ensemble and individual type-1 matrices can both exhibit saturation to values of $\Sigma^2(R,L)$ much smaller than \eref{rigidity} and have a spectral variance greatly exceeding \eref{rigidity}. Type-$M$ matrices, whose construction is detailed in the next section, exhibit similar behavior in $\Sigma^2(R,L)$ for small $M$, but we have not quantified how precisely $\Sigma^2(R,L)$ changes with increasing $M$.

Recent work by Prakash and Pandey\cite{Pandey2} shows that a two particle non-interacting embedded matrix ensemble\cite{Brody2} exhibits saturation of $\Sigma^2(L)$ on the ensemble average. Embedded matrix ensembles model the structure of many body systems by constructing eigenenergies out of random $k$-body interactions between $m$ particles, $k< m$. Ref.~\onlinecite{Pandey2} contains an extended discussion of saturation and helpful references. We do not pursue spectral variance further in this work.

\section{Statistics of integrable matrices of higher types}
\label{sec:HigherTypesResults}

\subsection{Ansatz type-M families}
\label{ss:AnsatzTypeM}

We do not yet have a method for directly generalizing the type-1 primary parametrization from Sect.~\ref{ss:Type1families} to higher type matrices that by definition have fewer commuting partners. Instead, we present another parametrization that produces a subset of integrable families of any type $M \ge 1$. The construction is in terms of $3N+1$ real parameters so that in choosing values for them one obtains a matrix $H(x) = xT + V$ with a desired number $n$ of nontrivial commuting partners ($n = N - M$) and no parameter-independent symmetries. As in the type-1 primary parametrization, the parameters can be traced back to eigenvalues of two commuting constant random matrices and a random vector.

Here we present the results; more details can be found in Ref.~\onlinecite{owusu1} while the rotationally invariant construction is given in Ref.~\onlinecite{ScYu}. Again in the diagonal basis of $V$, the most general member of an ansatz type-$M$ commuting family is
\beg
\begin{array}{l}
\dis \left[H \left(x\right)\right]_{ij}=x
\gamma_{i}\gamma_{j} \left(\dfrac{d_{i} -d_{j} }{\varepsilon_{i}-\varepsilon_{j}}\right)\dfrac{\Gamma_i +\Gamma_j}{2},\quad i\ne j,\\
\\
\dis\left[H \left(x\right)\right]_{ii}=\\
\dis \quad = d_{i} -  x\sum_{j\neq i} {\gamma_{j}^{2}
\left(\dfrac{d_{i} -d_{j} }{\varepsilon_{i}-\varepsilon_{j}}\right) \dfrac{1}{2} \dfrac{\left( \Gamma_i+\Gamma_j\right)\left( \Gamma_j+1\right)}{\Gamma_i+1}},\\
\end{array}
\label{Hmel2}
\en
where
\beg
\begin{array}{l}
\dis \braket{i|i}\equiv\sum_{j=1}^N\frac{\gamma_j^2}{(\lambda_i-\eps_j)^2}, \\
\\
\dis d_i=\frac{1}{x_0}\sum_{j=1}^{N-M}\frac{g_j}{\braket{j|j}}\frac{1}{\lambda_j-\eps_i}, \\
\\
\dis \Gamma_i =\pm \sqrt{1+\frac{1}{x_0}\sum_{j=N-M+1}^{N}\frac{P_j}{\braket{j|j}}\frac{1}{\lambda_j-\eps_i}}. 
\label{dm2}
\end{array}
\en
This parametrization gives all type-1, 2, and 3 integrable matrices and only a subset of such for higher types. We call matrices obtained by this construction \textit{ansatz type-M} as opposed to all type-$M$, these two notions being equivalent for $M=1,2,3$.

Basis-independent considerations from Ref.~\onlinecite{ScYu} identify $\lambda_i$ as eigenvalues of a matrix $\Lambda$ selected from the GOE and $\gamma_i$ as selected from a $\delta(1-|\gamma|^2)$ distribution, as was the case for the primary parametrization of type-1 matrices in Sect.~\ref{ss:Type1families}.  One may alternatively select the $\eps_i$ as eigenvalues of a GOE matrix $E$ and from them derive the $\lambda_i$. We find that this choice has no effect on the statistics. Unique to the ansatz parametrization are the ($N-M$) parameters $g_i$ and $M$ parameters $P_i$. Ref.~\onlinecite{ScYu} identifies these parameters as eigenvalues selected from an $N\times N$ GOE matrix $G$\cite{asterisk} satisfying $[G,\Lambda]=0$. The sign of $\Gamma_i$ can be chosen arbitrarily for each $i$ and each set of sign choices corresponds to a different commuting family.  The $\lambda_i$ by construction are solutions of the following equation with arbitrary (but fixed) real $x_0\ne0$:
\beg
\begin{split}
&f(\lam_i) \equiv\sum_{j=1}^{N}\dfrac{\gamma_{j}^2}{\lambda_i-\varepsilon_j}-\frac{1}{x_0}=0, \\
&F(\varepsilon_i) \equiv \sum_{j=1}^{N}\dfrac{1}{\braket{j | j}}\dfrac{1}{\lambda_j-\varepsilon_i}-x_0=0.
\end{split}
\label{Gaudin1}
\en
The second line of \eref{Gaudin1} follows from the first by writing both the partial fraction decomposition and factorized form of $F(z)=1/f(z)$ and matching residues. \esref{dm2} and \re{Gaudin1} mean that ansatz type-$M$ matrices are written in terms of an auxiliary primary type-1 problem with parameter $x_0$ and (unnormalized) eigenstates $|i\rangle$, see \eref{eq:selfConsist} and Ref.~\onlinecite{ScYu}. Note the important distinction between $x$ and $x_0$ -- namely that $x$ is free but $x_0$ is fixed for a given family of commuting matrices.

Due to the square root in the expression for $\Gamma_i$, \eref{dm2}, a given set of $P_i$ will typically result in a complex set of $\Gamma_i$. The matrix $H(x)$ will subsequently be complex symmetric, rather than real, although it will still satisfy all requirements of integrability. Because in this work we study the eigenvalues of real symmetric integrable matrices, we elect to reparametrize $\Gamma_i$ in a way that guarantees they be real without awkwardly scaling each set of $P_i$
\beg
\begin{array}{l}
\dis \Gamma_i =\pm \sqrt{\frac{\prod_{j=1}^N(\phi_j-\eps_i)}{\prod_{k=1}^N(\lambda_k-\eps_i)}},
\label{dm3}
\end{array}
\en
where the $M$ $\phi_j$ are real parameters such that (upon ordering $\eps_j$ and $\lambda_j$ for argument's sake) $\eps_j<\phi_j<\lambda_j$ if $x_0>0$ and $\lambda_j<\phi_j<\eps_j$ if $x_0<0$. The resulting $\Gamma_i$ are real-valued. As there is no existing basis-independent interpretation for $\phi_j$, we simply choose them from a uniform distribution on their allowed intervals. We find that the choice of $\phi_i$ or $P_i$ to generate the $\Gamma_j$ has a numerically undetectable effect on the eigenvalue statistics.

Varying parameters $g_j$ produces different matrices within the same commuting family, while varying the remaining parameters $\gamma_i,\lambda_i, \phi_i, x_0$ generates sets of matrices from different families. A natural way to choose a basis for the ansatz type-$M$ commuting family is to define the $n=N-M$ nontrivial $H^k(x)$ such that $g_j=\delta_{kj}$ in \eref{dm2} for  $1\le j\le N-M$. In other words,
\beg
\begin{split}
H^k(x)&=xT^k+V^k \mbox{ is given by Eq.$~\re{Hmel2}$ with} \\
d_i &\to d_i^k= \frac{1}{x_0}\dfrac{1}{\braket{k|k}}\dfrac{1}{\lambda_{k}-\varepsilon_i}.
\end{split}
 \label{him}
 \en
 for $k=1,\dots,N-M$. In particular,
 \beg
 \begin{split}
V^k &= \textrm{Diag}(d_1^k, d_2^k,\dots,d_N^k)
 \end{split}
 \label{vibasis}
 \en
A general member of the commuting family is 
\beg
H(x)=\sum_{k=1}^{N-M} g_k H^k(x).
\label{genedfin}
\en
up to a multiple of the identity trivial to the study of level spacing statistics.

Ansatz type-$M$ families have an exact solution in terms of a single equation similar to \eref{eq:selfConsist}  given in Ref.~\onlinecite{owusu1}, which has slight differences in notation as compared to here. To study level statistics of ansatz matrices, we numerically diagonalize them rather than use the computationally cumbersome exact solution.

A fundamental difference between ansatz type-$M$ matrices and the primary type-1 parametrization is that the eigenvalues of the matrix $V$ in the former are heavily constrained by \eref{dm2}, while in the latter they are free parameters. In particular, as explained  in Ref.~\onlinecite{ScYu} the primary type-1 $V$ is selected from the GOE, while  the ansatz $V$ is a certain primary type-1 matrix evaluated at $ x=-x_0$, i.e.,
\beg
V(x_0)=-x_0 T_{H_1} +H_1,
\label{genedfin2}
\en
where $H_1$ has $N-M$ arbitrary eigenvalues $g_i$ and $M$ eigenvalues equal to zero. By the results of Sect.~\ref{Type1}, ansatz $V=V(x_0)$ will typically have Poisson statistics. The resolution to this apparent disconnect between the two parametrizations is that for $|x_0| \ll 1$, $V(x_0)$ will have the eigenvalue statistics of $H_1$. We argue in Ref.~\onlinecite{ScYu} that the $N-M$ $g_i$ are a subset of eigenvalues of an $N\times N$ matrix from the GOE, so that for $M$ not too large and $x_0 \ll 1$ we obtain Wigner-Dyson statistics in ansatz $V$.

We then forgo studying crossovers in the coupling $x$ of level statistics of ansatz type-$M$ matrices $H(x)=xT +~V$ because ansatz $V$ have Poisson statistics for typical parameter choices. Instead, we focus on the behavior of the statistics with respect to parameter correlations, the number $M$ and the number of basis matrices. In all numerical work on ansatz matrices we set $x_0 = 1$, as this is a typical coupling value for the auxiliary type-1 problem.

\subsection{Correlations in ansatz parameters}
\label{ssec:correlationansatz}

\begin{figure}
\includegraphics[width=\linewidth]{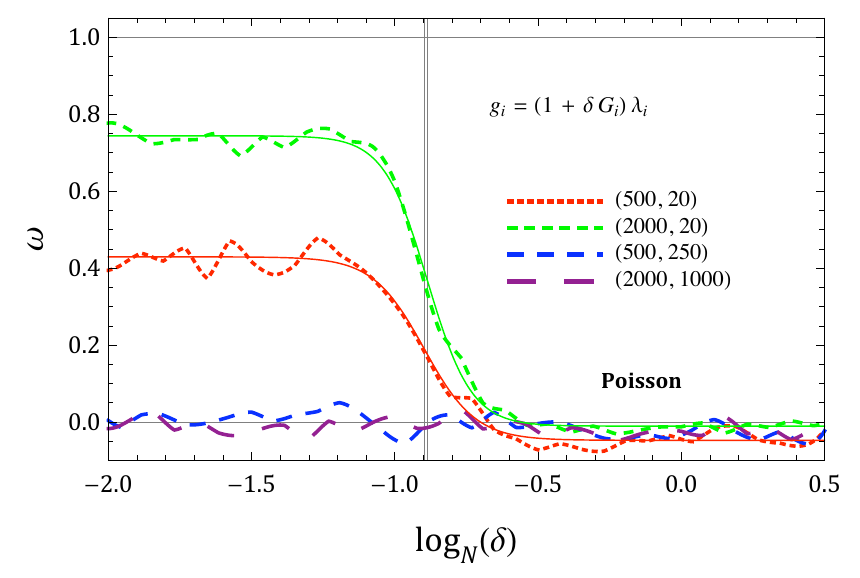}
\caption{(color online) Variation in the Brody parameter $\omega(\delta, N)$ when $g_i = \lambda_{i}(1 + \delta G_i)$ in the level statistics of $N\times N$ ansatz type-$M$ integrable matrices $H(x=1)$, \eref{Hmel2}, for various $N$ and $M$. Ordered pairs in the legend indicate size and type $(N,M)$ of the matrices, $\delta$  controls the strength of the perturbation from the point $\delta=0$ where the parameters $g_i$ and $\lambda_i$ defining these integrable matrices are correlated, and $G_i$ is an $\mathcal{O}(1)$ random number from a normal distribution. The crossover in $\delta$ for small $M$ is  similar to the primary type-1 crossovers in $\delta$ and $x$ seen in Figs.~\ref{fig:Type1crossoverx},\ref{fig:Type1CorrelationCrossover} and \ref{fig:Type1CorrelationCrossover2}.   For larger $M$, correlations cannot be introduced  by this method, see \eref{dm3}.  Despite type-$M$ matrices having fewer than the maximum number of conservation laws, the crossover still demonstrates the   scaling given in \eref{eq:BrodyFitFunctionDelta} (solid curves) with a crossover centered around $X_0 \sim -1$ (vertical line).  As before, deviations from correlation of size $\delta\propto N^{-0.5}$ are enough for the statistics to  become Poisson. Each plotted value $\omega(\delta,N)$ is computed for the combined level spacing distribution of several matrices from the ensemble. For the case of correlations in ansatz matrices, we choose all $\Gamma_k > 0$ in order to avoid pathological statistics in $H(x)$.}
 \label{fig:Type20CorrN500LevSta}
\end{figure}

Building on the results of Sect.~\ref{ssec:correlation1}, here we explore effects of parameter correlations on the statistics in general type-$M$ ansatz matrices. Introducing correlations between $d_i$ and $\varepsilon_i$ in this case is more complicated than in Sect.~\ref{ssec:correlation1} because the $d_i$ here are not all independent. Fortunately, Eq.~(\ref{dm2}) admits a simple way to produce such correlations. As an example, consider the case when $g_j = \lambda_{j}$
\beg
\begin{array}{l}
\dis d_i = \frac{1}{x_0} \sum_{j=1}^{N-M}\dfrac{\lambda_{j} -\varepsilon_i + \varepsilon_i}{\braket{j | j}}\dfrac{1}{\lambda_{j}-\varepsilon_i}\\
\\
\dis \quad =  \frac{1}{x_0} \,\varepsilon_i \left(\sum_{j=1}^{N-M}\dfrac{1}{\braket{j | j}}\dfrac{1}{\lambda_{j}-\varepsilon_i}\right)  + (\textrm{const}) \\
\\
\dis \quad = \varepsilon_i\left(1 - \frac{1}{x_0}\sum_{j=1}^{M}\dfrac{1}{\braket{j|j}}\dfrac{1}{\lambda_{j}-\varepsilon_i}\right) + (\textrm{const}),
\label{dm3}
\end{array}
\en
where the second part of Eq.~(\ref{Gaudin1}) was used. The sums in the third line of Eq.~(\ref{dm3}) introduce a randomizing factor that has a weak effect for small  $M$ but that destroys the correlation between $d_i$ and $\varepsilon_i$ at intermediate values of $M$. Fig.~\ref{fig:Type20CorrN500LevSta} shows the now familiar level statistics crossover in $\delta$ for ansatz matrices of different size and type with $g_k = \lambda_{k}(1+\delta G_k)$, where $G_k$ is an $\mathcal{O}(1)$ random number chosen from a normal distribution and $\delta$ a parameter controlling the size of the perturbation. Just as in Sect.~\ref{ssec:correlation1}, the crossover to Poisson statistics is centered about $\delta \sim N^{-1}$. More generally, we can induce level repulsion in ansatz type-$M$ matrices if $M \ll N$ when $g_k = f(\lambda_{k})$, a smooth function of $\lambda_{k}$.

\subsection{Basis matrices: ansatz higher types}
\label{ssec:basisansatz}

We now generalize the type-1 results of Sect.~\ref{ssec:basismatrices1} to apply to all ansatz type-$M$ matrices. Recall that a general ansatz type-$M$ matrix $H(x) = xT + V$ can be written as a linear combination of basis matrices $H^k(x)$ for which $g_i = \delta_{ik}$ (see \eref{genedfin}).

\begin{figure}
\includegraphics[width=\linewidth]{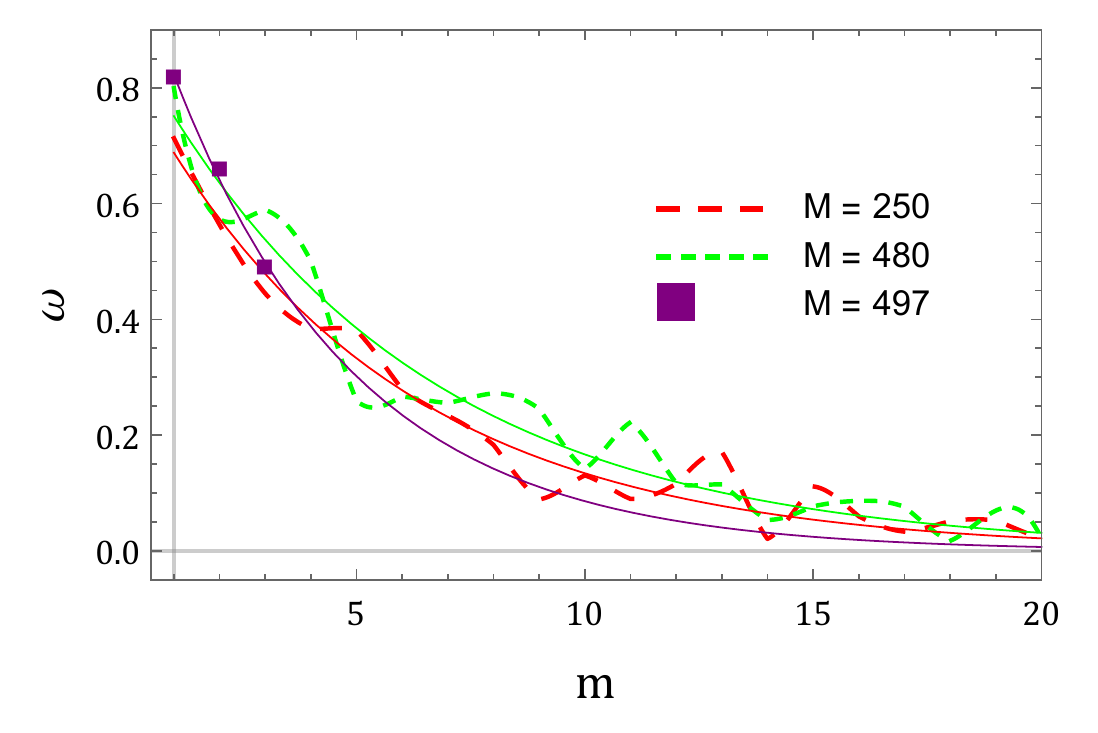}
\caption{(color online) Graph of the Brody parameter $\omega(m,N)$ given by \eref{eq:Brody} vs. number $m$ of ansatz type-$M$ basis matrices $H^k(x)$, see \eref{him}, contained in linear combination $H(x) = \sum_{k=1}^m g_k H^k(x)$ for $N = 500$, $x=1$. The fits presume exponential decay and are expressed in terms of two parameters $(a,b)$ from Eq.~(\ref{eq:basisfit}). For $M = (250, 480)$ we find the decay constant $b =(1.13, 1.04)$, indicating that we only need $m_{\textrm{min}} \approx \log{N}$ conservation laws for Poisson statistics to emerge,  independent of type. We do not observe Poisson statistics for $M = 497$ because the maximum number of nontrivial basis matrices is 3 in this case, and we see that we need at least $\sim 15$ conservation laws for Poisson statistics to start emerging for $N = 500$. See Fig.~\ref{AnsatzBasis2000} for a similar plot for $N = 2000$ and Fig.~\ref{fig:Type1Basis} for the same concept in type-1 matrices. Each plotted value $\omega(m,N)$ is computed for the combined level spacing distribution of several matrices from the ensemble.}
 \label{AnsatzBasis500}
\end{figure}

\begin{figure}
\includegraphics[width=\linewidth]{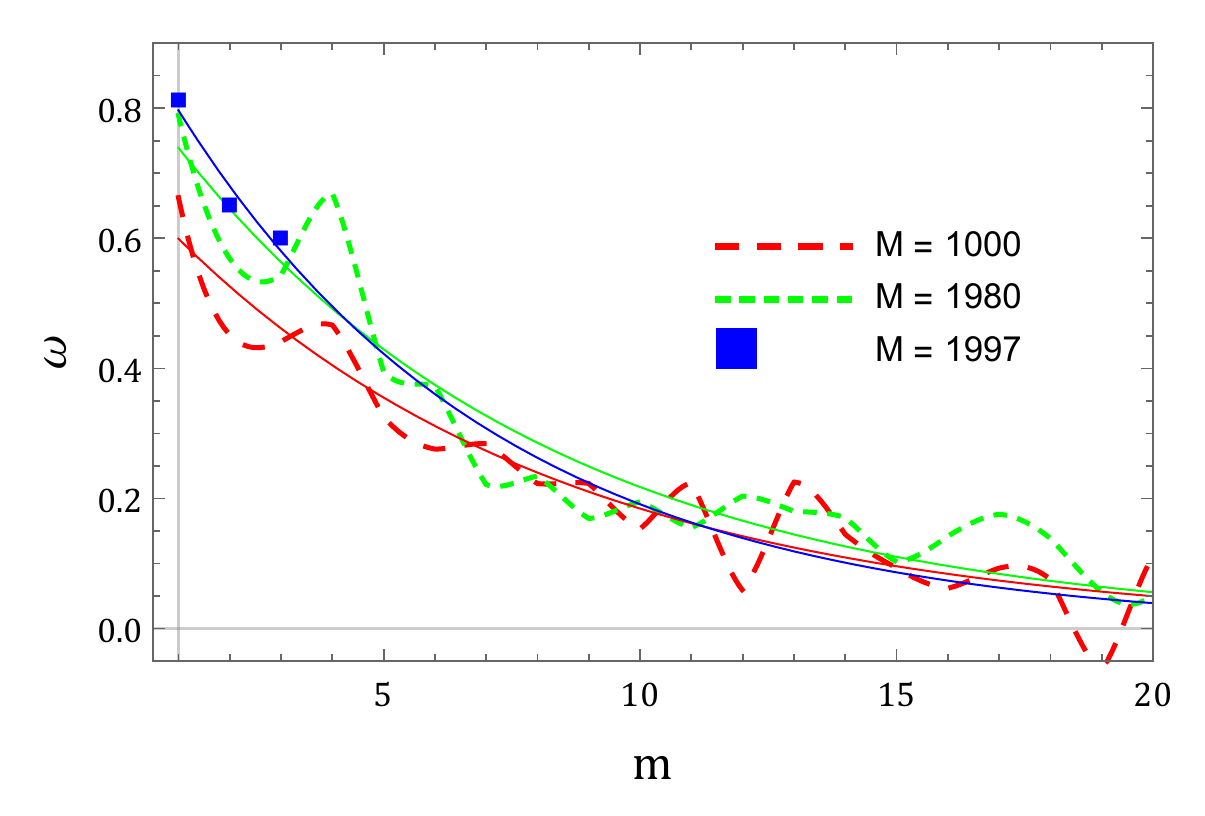}
\caption{(color online)  Brody parameter $\omega(m,N)$ (see \eref{eq:Brody}) vs. number $m$ of ansatz type-$M$ basis matrices $H^k(x)$, see \eref{him}, contained in linear combination $H(x) = \sum_{k=1}^m g_k H^k(x)$ for $N = 2000$, $x=1$. The fits presume exponential decay and are expressed in terms of two parameters $(a,b)$ from Eq.~(\ref{eq:basisfit}). For $M = (1000, 1980)$ we find the decay constant $b =(0.99,1.03)$, indicating that we only need $m_{\textrm{min}} \approx \log{N}$ conservation laws for Poisson statistics to emerge, independent of type. We do not observe Poisson statistics for $M = 1997$ because the maximum number of nontrivial basis matrices is 3 in this case, and we see that we need at least $\sim 20$ conservation laws for Poisson statistics to start emerging for $N = 2000$. See Fig.~\ref{AnsatzBasis500} for a similar plot for $N = 500$ and Fig.~\ref{fig:Type1Basis} for the same concept in type-1 matrices. Each plotted value $\omega(m,N)$ is computed for the combined level spacing distribution of several matrices from the ensemble.}
 \label{AnsatzBasis2000}
\end{figure}

We see again in Figs.~\ref{AnsatzBasis500} and \ref{AnsatzBasis2000} that Poisson statistics emerge for relatively small linear combinations of basis matrices. Denoting $m$ as the number of conservation laws contained in a linear combination, i.e.,
\beg
H(x)=\sum_{i=1}^m g_k H^k(x),\quad m\le N-M,
\en
we investigate the Brody parameter $\omega(m,N)$ from \eref{eq:basisfit}. In Fig.~\ref{AnsatzBasis500}, $N = 500$, $\omega(m,N)$ decays to zero as a function of $m$ in nearly the same way for $M = 470$ as for $M = 20$. It is only for very large $M$, such as $M = 497$, that level clustering is forbidden, and this only because we can use a maximum of 3 nontrivial basis matrices. Similar behavior emerges for $N = 2000$ in Fig.~\ref{AnsatzBasis2000}. For all $N$ and $M$ tested we find $b \sim 1$ (with precise values given in the captions). Therefore, we can estimate a similar bound as in Sect.~\ref{ssec:basismatrices1} for the minimum number of conservation laws needed for Poisson level statistics, namely $m_{\textrm{min}} < \mathcal{O}(N^{\alpha})$ where $0 < \alpha < 0.25$, obtained from the $M= N/2$ cases. Since $m$ cannot exceed the total number of conservation laws $n=N-M$ for type-$M$ matrices, this provides a lower bound $n_{\textrm{min}}=m_{\textrm{min}} < \mathcal{O}(N^{\alpha})$ consistent with $m_{\textrm{min}}\approx \log N$.

\section{\label{exception}Analytical results: perturbation theory}

Some of the numerical observations found in Sects.~\ref{Type1} and \ref{sec:HigherTypesResults} can be understood using perturbation theory in the parameter $x$. We restrict our analysis to the primary type-1 parametrization because our arguments for this case are much more transparent than for the ansatz construction. The analysis for ansatz matrices is similar.

The eigenvalues $\eta_m(x)$ of $H(x)$ to first order in $x$ are given by the second equation in \eref{eq:HType1Param}, where we set constant $|\gamma_j|^2=N^{-1}$ for clarity and to achieve proper scaling for large $N$
\beg
 \eta_m(x)\approx d_{m} -\dfrac{x}{N}\sum_{j\neq m} 
\left(\dfrac{d_{m} -d_{j} }{\varepsilon_{m}-\varepsilon_{j}}\right).
\label{pert}
\en
The first term comes from $V$, which has a Wigner-Dyson $P(s)$, and the second term from $T$, which is determined by the integrability condition and whose level statistics we do not control. Let us estimate the $x$ at which the two terms in \eref{pert} become comparable.  Note that $d_k$ and $\eps_k$ both lie on $\mathcal{O}(1)$ intervals so that $T$ and $V$ scale in the same way for large $N$. Suppose $\eps_k$ are ordered as $\eps_1<\eps_2<\dots <\eps_N$. When $d_k$ and $\eps_k$ are uncorrelated $d_m-d_j$ is  $\mathcal{O}(1)$ when $j$ is close to $m$, i.e., when $(\eps_m-\eps_j)=\mathcal{O}(N^{-1})$. The second term in \eref{pert} is then $x c_m \ln N$, where $c_m= \mathcal{O}(1)$ is a random number only weakly correlated with $d_m$. We performed simple numerical tests that confirm this scaling argument.

If we now order $d_m$, $c_m$ in general will not be ordered, i.e., if $d_{m+1}>d_m$ is the closest level to $d_m$ and therefore $(d_{m+1}-d_m)= \mathcal{O}(N^{-1})$, the corresponding difference $(c_{m+1} -c_m)= \mathcal{O}(1)$. The contributions to level-spacings from the two terms in \eref{pert} become comparable for $x=x_c\approx 1/(N\ln N)$. It makes sense that the second term introduces a trend towards a Poisson distribution because it is a (nonlinear) superposition of $\eps_k$ and $d_k$ -- eigenvalues of two uncorrelated random matrices. Thus, we expect a crossover from Wigner-Dyson to Poisson statistics near $x=x_c$. In our numerics we observe a crossover over the range $N^{-1.5} \lesssim x \lesssim N^{-0.5}$ centered about $x_c \sim N^{-1}$ likely because we do not reach large enough $N$ to detect the log component of the crossover.

This argument breaks down when $d_k=f(\eps_k)$, since in this case $(d_m-d_j)= \mathcal{O}(N^{-1})$ when $(\eps_m-\eps_j)= \mathcal{O}(N^{-1})$. The two terms in \eref{pert} become comparable only at $x= \mathcal{O}(1)$; moreover, the second term no longer trends towards Poisson statistics. Relaxing the correlation between $d_k$ and $\eps_k$ with $d_k=f(\eps_k)(1+\delta D_k)$, $D_k = \mathcal{O}(1)$, and going through the same argument, one expects a  crossover to Poisson statistics at $\delta =  \mathcal{O}(1/N\ln N)$ when $x= \mathcal{O}(1)$.

The level repulsion observed in basis matrices is a consequence of the level repulsion implicit in the parameters $\lambda_i$, independent of the choice of $\eps_i$, see the text below \eref{BCS} and Fig.~\ref{diffparamchoices}. Indeed, basis matrices $H^i(x)$ in the primary type-1 parametrization, Eq.~(\ref{eq:HType1Param}), have eigenvalues $\eta^i_j(x) =x \gamma_i^2(\lambda_j - \eps_i)^{-1}$, which is a smooth function of $\lambda_j$ except near $\eps_i$. The $\eta^i_j(x)$ therefore inherit the level repulsion of the $\lambda_j$.   Analogous reasoning applies to ansatz basis matrices.

\section{Ergodicity in integrable matrix ensembles}
\label{sec:ergodicity}

The discussion and figures in this section make frequent reference to the ``primary" construction of type-1 integrable matrices and the ``ansatz" construction of type-$M$ integrable matrices. These parametrizations are introduced in Sect.~\ref{ss:Type1families} and Sect.~\ref{ss:AnsatzTypeM}, respectively. Ensemble averages are taken with respect to the probability distributions for integrable matrices introduced in Ref.~\onlinecite{ScYu}.

One of the goals of this work is to determine the extent to which ensembles of integrable matrices are ``ergodic.'' Intuitively, an ensemble is called ergodic if a single randomly selected member has properties that are typical of the entire ensemble. Bohigas and Gianonni\cite{bohigas} expound the subject in generality for random matrices, and here we focus numerically on the meaning of ergodicity with regards to the nearest-neighbor level spacing distribution of integrable matrices. Rigorous results on ergodicty for Gaussian ensembles and the Poisson ensemble were derived by Pandey\cite{Pandey1}.

We distinguish between three separate ways of generating nearest-neighbor eigenvalue spacing distributions for $N\times N$ integrable matrix ensembles. We call $P_{i,N,R}(s)$ the level spacing distribution, normalized to unity, of the $i$-th member of the ensemble obtained from a spectral region $R$ containing many eigenvalues (infinitely many as $N \to \infty$). The normalized distribution of spacings in $R$ from all matrices in the ensemble is called $\mathcal{P}_{N,R}(s)$. A third way to characterize spacing statistics is through the normalized distribution of the $j$-th eigenvalue spacing of all matrices in the ensemble, which we call $p_{N,j}(s)$. Both the regions $R$ and the numbers $j$ are stipulated to be far from the edges of the spectrum. In general, $P_{i,N,R}(s)$, $\mathcal{P}_{N,R}(s)$ and $p_{N,j}(s)$ are distinct distributions. Conceptually, $\mathcal{P}_{N,R}(s)$ and $p_{N,j}(s)$ are ensemble properties while $P_{i,N,R}(s)$ characterizes the spectrum of an individual matrix. In the following definitions, we assume that the spacing distributions converge to a well-defined limit as $N\to \infty$, unlike known pathological examples such as the semiclassical spacing distribution of a harmonic chain\cite{jberry}. This assumption is supported numerically.

\begin{figure}
\includegraphics[width=\linewidth]{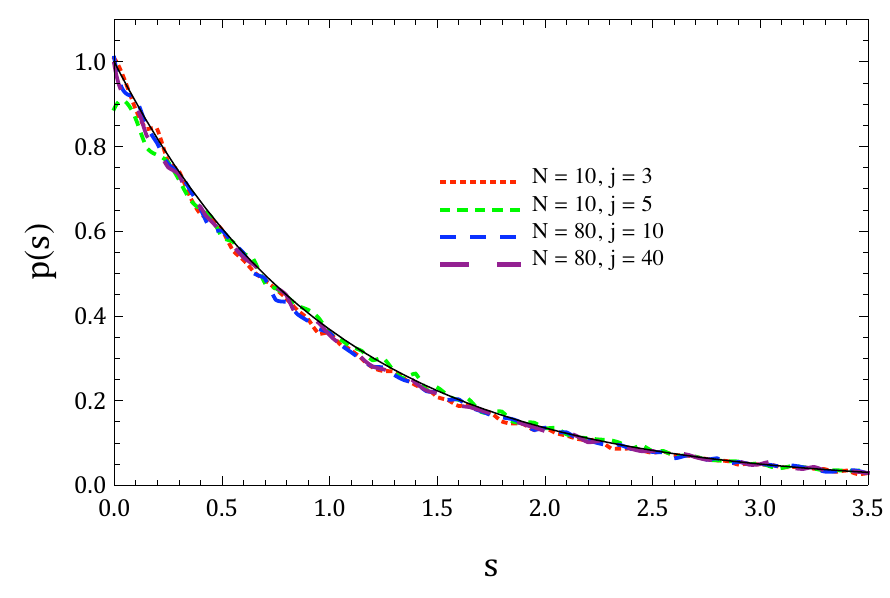}
\caption{(color online) Demonstrating the stationary property \eref{eq:SErgo} in type-1 $N\times N$ matrices $H(x)$, $x=1$ in the primary parametrization. The four numerical curves show the statistics $p_{N,j}(s)$ for $(N,j)$ = $(10,3)$, $(10,5)$, $(80,10)$ and $(80,40)$, each containing $10^5$ eigenvalue spacings. The statistics are nearly independent of $j$ for $N = 10$, and for $N = 80$ there is no perceptible difference between $j = 10$ and $j = 40$. The solid line is a Poisson distribution $p(s) = e^{-s}$. Stationarity is shown to hold also for type-$M$ ansatz matrices in Fig.~\ref{HorzErgAnsatz}.}
\label{HorzErgN10N80}
\end{figure}

\begin{figure}
\includegraphics[width=\linewidth]{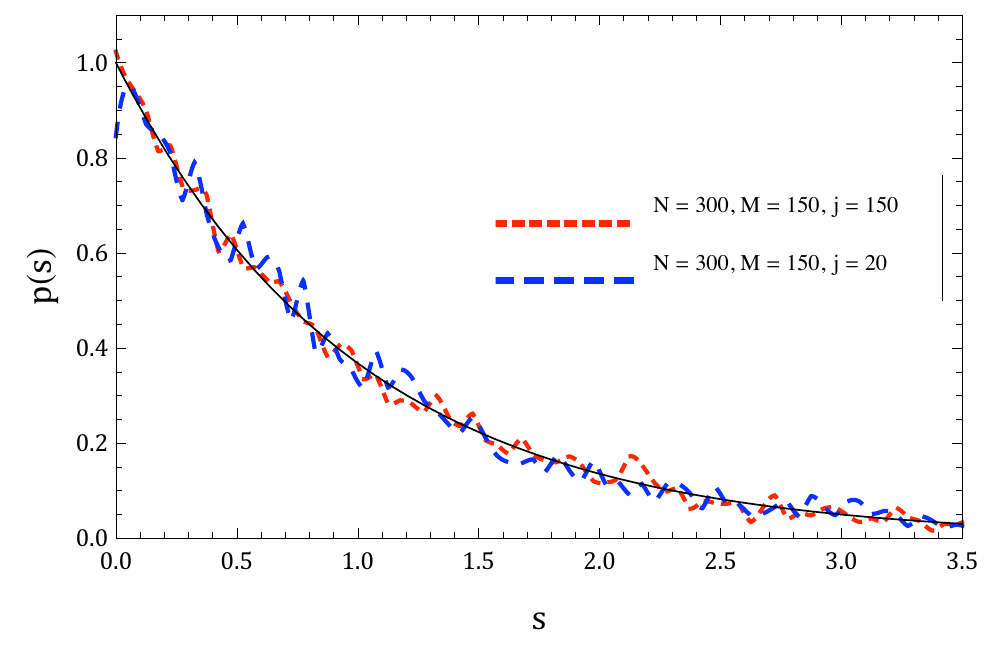}
\caption{(color online) Demonstrating the stationarity property \eref{eq:SErgo} in ansatz type-150 $N\times N$ matrices $H(x)$, $x=1$ and $N = 300$. The two numerical curves show the statistics $p_{N,j}(s)$ for $(N,j)$ = $(300,150)$ and $(300,20)$, each containing $\sim 10^4$ eigenvalue spacings. The statistics are nearly independent of $j$, although higher $N$ would be needed in order for the differences to disappear. The solid line is a Poisson distribution $p(s) = e^{-s}$.}
\label{HorzErgAnsatz}
\end{figure}

We now describe a precise notion\cite{Pandey1} of ergodicity that characterizes the limiting behavior of $P_{i,N,R}(s)$, $\mathcal{P}_{N,R}(s)$ and $p_{N,j}(s)$ as $N \to \infty$. First, we must determine whether $p_{N,j}(s)$ is asymptotically stationary, i.e., independent of $j$
\begin{equation}
\begin{array}{l}
\dis \lim_{N \to \infty} p_{N,j}(s) = p(s).
\end{array}
\label{eq:SErgo}
\end{equation}
In the case of type-1 matrices in the primary parametrization, we see in Fig.~\ref{HorzErgN10N80} that the graphs of two different $p_{10,j}(s)$ closely resemble those of two different $p_{80,j}(s)$, the latter of which are clearly Poisson. The same is true for ansatz matrices of any type, but the convergence to a Poisson distribution does not become apparent until $N = 300$ as in Fig.~\ref{HorzErgAnsatz}. We conclude that \eref{eq:SErgo} is true for integrable matrices. 

\begin{figure}
\includegraphics[width=\linewidth]{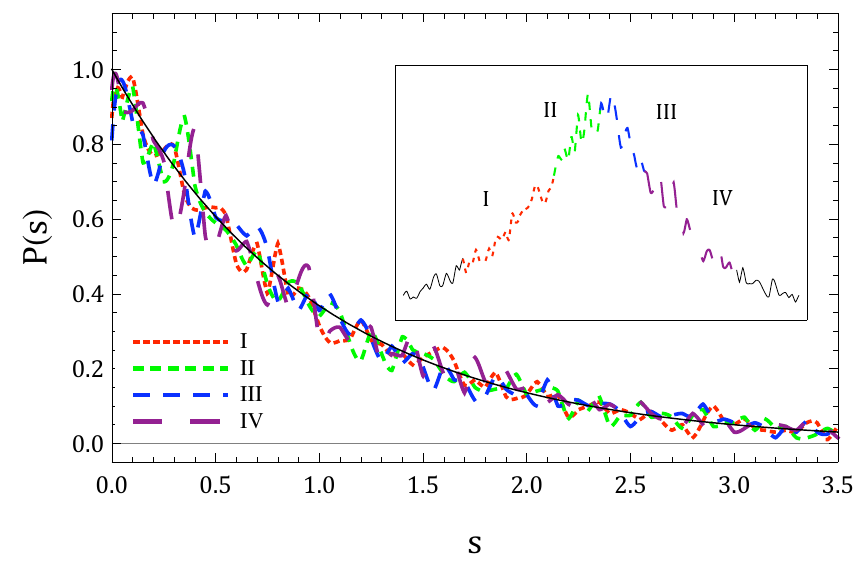}
\caption{(color online) Demonstrating spectral stationarity \eref{eq:SErgo35} in type-1 matrices. Shown are the level statistics $P_{i,N,R}(s)$ of a single ($i$-th member of the ensemble) type-1 integrable matrix $H(x)$, $x=1$ and $N = 20000$, for different regions $R$ of its spectrum containing 4000 eigenvalues each. The inset shows the density of states of this matrix and indicates which numerical curve corresponds to which region $R$. The distributions $P_{i,N,R}(s)$ shown are independent of $R$, indicating that type-1 matrix spectra are stationary with respect to nearest neighbor level statistics.  Noting that these distributions are Poisson, $P_{i,N,R}(s) \approx e^{-s}$ (solid curve) and comparing to Fig.~\ref{100matricesN2000M1} which gives $\mathcal{P}_{N',R}(s) \approx e^{-s}$ for $N' = 2000$, we see that ergodicity, \eref{eq:SErgo3}, is satisfied for type-1 integrable matrices.}
\label{1matrixN20000M1}
\end{figure}

\begin{figure}
\includegraphics[width=\linewidth]{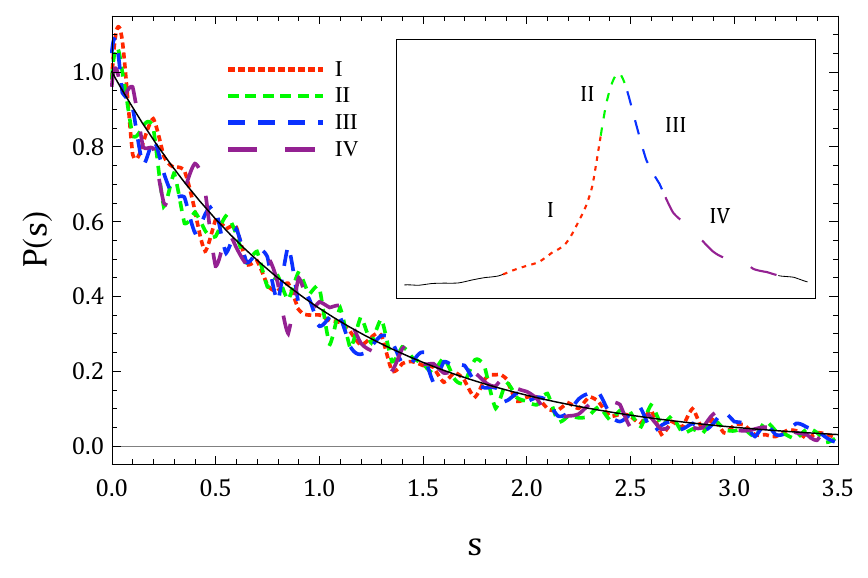}
\caption{(color online) Level statistics $P_{i,N,R}(s)$ of a single integrable matrix $H(x)$, $x=1$, $N = 20000$ and $M = 10000$, for different regions $R$ of its spectrum (the subscript $i$ indicates $H(x)$ is the $i$-th matrix in the ensemble) containing 4000 eigenvalues each. Inset: the density of states of $H(x)$  showing the correspondence between the distributions and regions $R$. The distributions $P_{i,N,R}(s)$  are independent of $R$, indicating that type-$M$ matrix spectra are stationary with respect to nearest neighbor level statistics, i.e., \eref{eq:SErgo35} holds. Noting that these distributions are Poisson, $P_{i,N,R}(s) \approx e^{-s}$ (solid curve) and comparing to Fig.~\ref{100matricesN2000M1000} which gives $\mathcal{P}_{N',R}(s) \approx e^{-s}$ for $N' = 2000$ $M' = 1000$, we verify the ergodic property, \eref{eq:SErgo3}.}
\label{1matrixN20000M10000}
\end{figure}

\begin{figure}
\includegraphics[width=\linewidth]{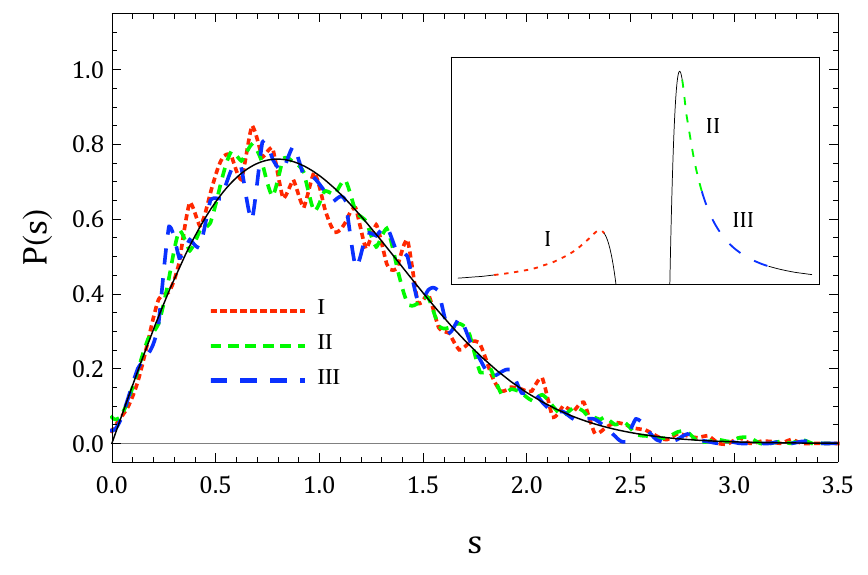}
\caption{(color online) Demonstrating spectral stationarity, \eref{eq:SErgo35}, in level statistics of primary type-1 basis matrices (defined in Sect.~\ref{ss:Type1families}). Shown are the level statistics $P_{i,N,R}(s)$ of a single type-1 integrable basis matrix, $x=1$ and $N = 20000$, for different regions $R$ of its spectrum (the subscript $i$ indicates $H(x)$ is the $i$-th matrix in the ensemble). Each spectral region $R$ contains $4000$ eigenvalues. The inset shows the density of states of this matrix and indicates which numerical curve corresponds to which region $R$. The distributions $P_{i,N,R}(s)$ shown are independent of $R$, indicating that type-1 basis matrix spectra are stationary with respect to level statistics. Even though there is a band gap, the level statistics on either side of the gap are the same. The solid curve is the Wigner surmise $P(s) = \frac{\pi}{2}s e^{-\frac{\pi}{4}s^2}$.}
\label{1matrixN20000M1basis}
\end{figure}

\begin{figure}
\includegraphics[width=\linewidth]{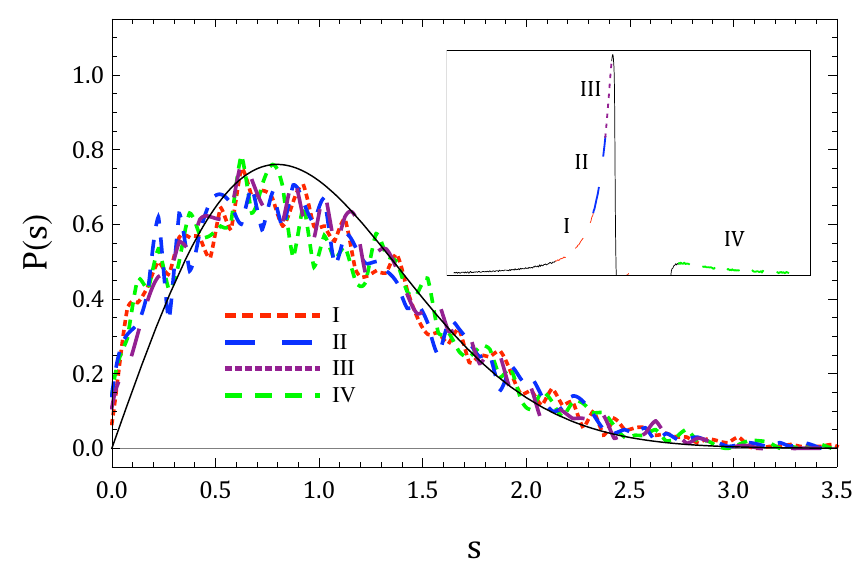}
\caption{(color online) Demonstrating spectral stationarity, \eref{eq:SErgo35} in level statistics of ansatz basis matrices (defined in Sect.~\ref{ss:AnsatzTypeM}). Shown are the level statistics $P_{i,N,R}(s)$ of a single type-10000 integrable ansatz basis matrix, $x=1$ and $N = 20000$, for different regions $R$ of its spectrum (the subscript $i$ indicates $H(x)$ is the $i$-th matrix in the ensemble). The inset shows the density of states of this matrix and indicates which numerical curve corresponds to which region $R$. The distributions $P_{i,N,R}(s)$ shown are independent of $R$, indicating that type-$M$ basis matrix spectra are stationary with respect to level statistics. Even though there is a band gap, the level statistics on either side of the gap are the same. The solid curve is the Wigner surmise  $P(s) = \frac{\pi}{2}s e^{-\frac{\pi}{4}s^2}$. Regions I - III use 4000 eigenvalues apiece, while region IV uses only 3000 and gets to within 1\% of the spectrums edge.}
\label{ansatz basis}
\end{figure}

We now turn to the notion of spectral averaging, i.e., the function $P_{i,N,R}(s)$. If \eref{eq:SErgo} holds, the ensemble averaged $P_{i,N,R}(s)$, called $\mathcal{P}_{N,R}(s)$, satisfies
\begin{equation}
\begin{array}{l}
\dis \lim_{N \to \infty} \mathcal{P}_{N,R}(s) = p(s),
\end{array}
\label{eq:SErgo2}
\end{equation}
independent of the region $R$. In practice, we numerically unfold the spectrum (see Appendix~\ref{appA}) in order to take into account any effects a non-stationary mean level spacing can have on $P_{i,N,R}(s)$, which characterizes fluctuations about the mean level spacing. In this work, we say integrable matrices are spectrally stationary if
\begin{equation}
\begin{array}{l}
\dis \lim_{N \to \infty} P_{i,N,R}(s) = P_i(s),
\end{array}
\label{eq:SErgo35}
\end{equation}
and ergodic with respect to nearest neighbor level statistics if
\begin{equation}
\begin{array}{l}
P_i(s) = p(s).
\end{array}
\label{eq:SErgo3}
\end{equation}
Two points are to be made about \eref{eq:SErgo35} and \eref{eq:SErgo3}. First, \eref{eq:SErgo35} is similar in spirit to, but not implied by, \eref{eq:SErgo}. Figs.~\ref{1matrixN20000M1}-\ref{ansatz basis} show for various integrable matrices, basis matrices included, that the level statistics from a single large matrix, $P_{i,N,R}(s)$, do not depend on the spectral region $R$ used.


\begin{figure}
\includegraphics[width=\linewidth]{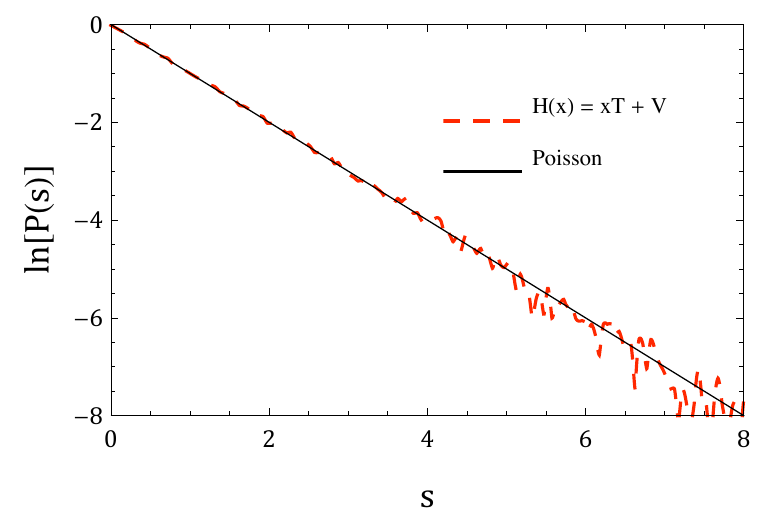}
\caption{(color online) Demonstrating ergodicity \eref{eq:SErgo3} in  type-1 matrices (continuing from Fig.~\ref{1matrixN20000M1}). A plot of $\ln{\mathcal{P}_{N,R}(s)}$, $N = 2000$ for 100 type-1 matrices in the primary parametrization. We do not specify the spectral region $R$ because Fig.~\ref{1matrixN20000M1} shows that the statistics are independent of $R$. The solid line is $f(s) = -s$, indicating that $\mathcal{P}_{N,R}(s)$ is indeed Poisson for $N = 2000$ type-1 matrices.}
\label{100matricesN2000M1}
\end{figure}

\begin{figure}
\includegraphics[width=\linewidth]{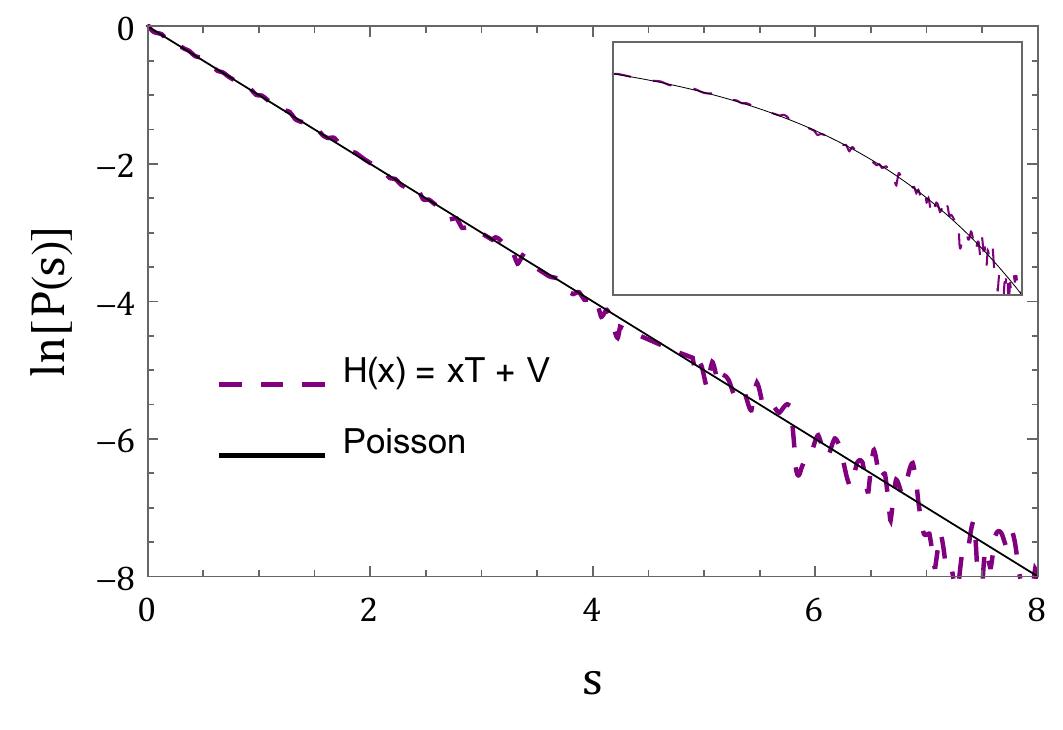}
\caption{(color online) Demonstrating ergodicity \eref{eq:SErgo3} in type-$M$ ansatz matrices (continuing from Fig.~\ref{1matrixN20000M10000}). A plot of $\ln{\mathcal{P}_{N,R}(s)}$, $N = 2000$ for 100 type $M =1000$ matrices in the ansatz parametrization. We do not specify the spectral region $R$ because Fig.~\ref{1matrixN20000M10000} shows that the statistics are independent of $R$. The solid line is $f(s) = -s$, indicating that $\mathcal{P}_{N,R}(s)$ is indeed Poisson for $N = 2000$ type-1000 matrices. Inset: Log-log plot of the same data.}
\label{100matricesN2000M1000}
\end{figure}

Second, the limiting distribution is independent of the index $i$, which means that a single matrix's spacing distribution is typical of the ensemble. In rigorous work on Gaussian ensembles\cite{Pandey1}, this is proved by showing the ensemble averaged variance of $ P_{i,N,R}(s)$ vanishes as $N\to \infty$. In this work, we compare numerically generated graphs of spectral spacing distributions to ensemble averaged ones for large $N$. By comparing  Figs.~\ref{100matricesN2000M1}, \ref{100matricesN2000M1000} to Figs.~\ref{1matrixN20000M1}, \ref{1matrixN20000M10000}, we see that for large $N$, $P_{i,N,R}(s) \to p(s)$.

The properties of stationarity and ergodicity are useful if they set in quickly for small $N$, because smaller matrices are more accessible both analytically and computationally. A classic example in Gaussian random matrix theory is the Wigner surmise, derived from $2\times 2$ matrices (see Fig.~\ref{fig:WigSurmise}), which is extremely useful for characterizing $p(s)$ in the GOE.

\begin{figure}
\includegraphics[width=\linewidth]{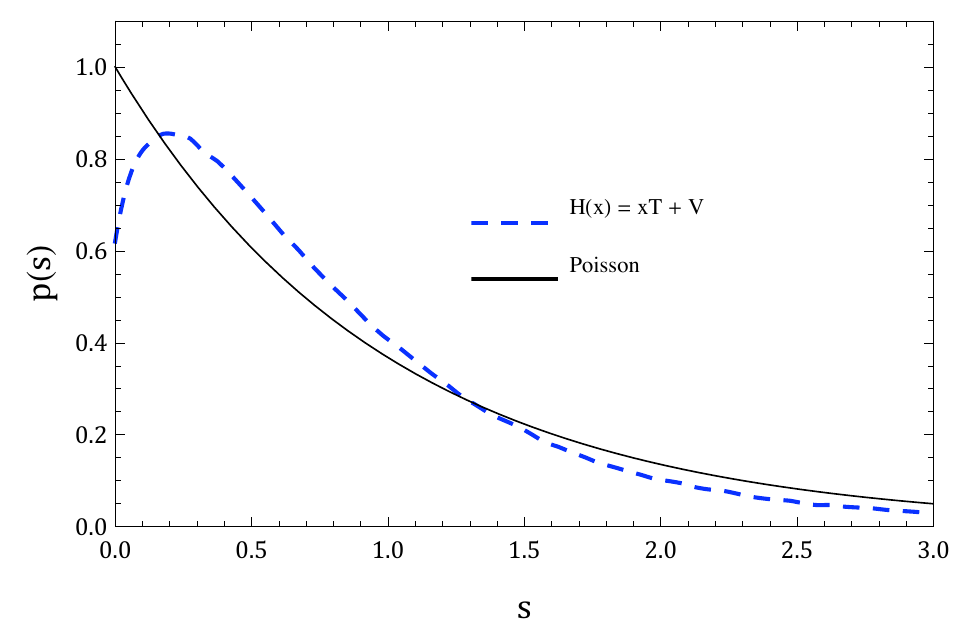}
\caption{(color online) Plot of the statistics $p_{3,2}(s)$, the second spacing of $10^6$ type-1 integrable matrices $H(x)$ of size $N = 3$ with $x = 1$. The distribution is not Poisson (solid line) and actually has a power law tail (see Fig.~\ref{fig:Type1N3loglog} for more on the tail). In order to observe the limit $p(s)$ of type-1 integrable matrices, defined in \eref{eq:SErgo}, we need to go to larger $N$ as in Fig.~\ref{HorzErgN10N80}.}
\label{fig:Type1N3}
\end{figure}

\begin{figure}
\includegraphics[width=\linewidth]{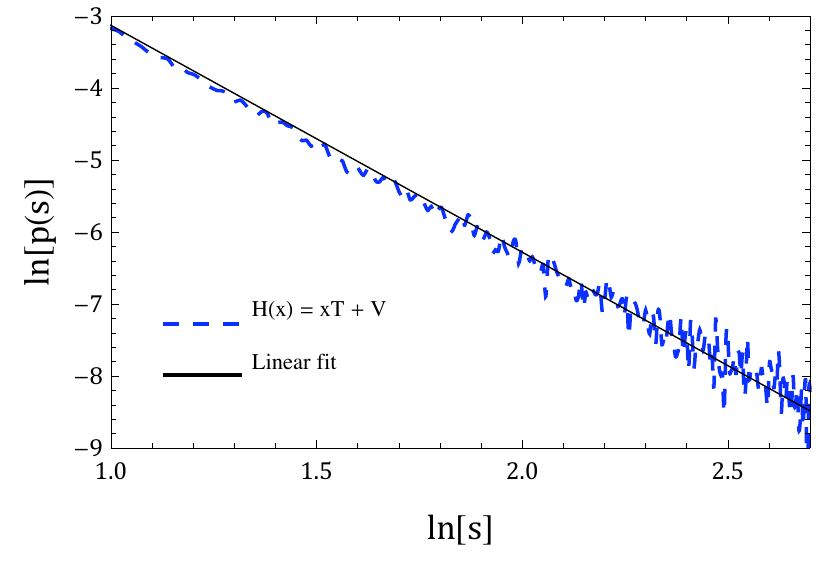}
\caption{(color online) Log-log plot of the tail of the distribution $p_{3,2}(s)$ shown in Fig.~\ref{fig:Type1N3}, the statistics of the second spacing of $10^6$ primary type-1 integrable matrices $H(x)$ of size $N = 3$ with $x = 1$. The linear fit $f(z) = -3.15z+0.02$ shows that this portion of the tail of the distribution $p_{3,2}(s)$ follows a power law $s^{-\alpha}$ with exponent $\alpha \approx 3.15$. Because the distribution $p_{N,j}(s)$ transitions to Poisson for large $N$, as evidenced by Fig.~\ref{HorzErgN10N80} for type-1 primary matrices and Fig.~\ref{HorzErgAnsatz} for type-$M$ ansatz matrices, we conclude that exponential behavior in the far tail of $p_{N,j}(s)$  likely emerges only in the limit $N\to \infty$.}
 \label{fig:Type1N3loglog}
\end{figure}

We have seen that the nearest neighbor level statistics of integrable matrices $H(x)$ are generally stationary and ergodic, but the property does not set in for small $N$ as quickly as it does for Gaussian random matrices. As an example, Figs.~\ref{fig:Type1N3}, \ref{fig:Type1N3loglog} show $p_{3,2}(s)$, the distribution of the second eigenvalue spacing for $N = 3$, $M = 1$. This distribution differs markedly from a Poisson distribution, especially in the small $s$ and large $s$ regions. For small $s$ there is slight level repulsion and for large $s$ Fig.~\ref{fig:Type1N3loglog} shows that the decay of $p_{3,2}(s)$ is a power law. Numerical data generated in Sects.~\ref{Type1} and \ref{sec:HigherTypesResults} used both $P_{i,N,R}(s)$ and $\mathcal{P}_{N,R}(s)$ to represent level statistics of integrable matrices. The results of this section show that for large $N$, it is valid to treat these two distinct distributions as equal.

\section{Conclusion}

Just as ensemble averages in ordinary RMT are used to predict the average behavior of generic quantum systems, there now exists an analogous ensemble theory- integrable matrix theory - which we have used to firmly establish the source of Poisson statistics and exceptions in quantum integrable models.

The goal of this work was to demonstrate two properties of ensembles of type-$M$ integrable $N\times N$ matrices linear in a coupling parameter $H(x) = xT + V$ as $N\to \infty$:

1) The nearest neighbor spacing distribution $P(s)$ is Poisson, $P(s) = e^{-s}$, for generic choices of parameters for almost all $M$. There are cases of level repulsion, but they correspond to sets of measure zero in parameter space.

2) Integrable matrix ensembles are both stationary and ergodic with respect to nearest neighbor level statistics as defined in Sect.~\ref{sec:ergodicity}. It remains to show whether this ergodicity extends to longer range spectral statistics, such as the number variance $\Sigma^2(L)$.

We find that integrable $N\times N$ matrices $H(x)$ have Poisson statistics as long as the number of conservation laws exceeds $n_{\textrm{min}} \approx \log{N}$   (or at most $n_{\textrm{min}} < N^{0.25}$).  Basis-independent considerations require (for type-1) GOE statistics at a fixed $x_0$, but we find that Poisson statistics return at deviations $\delta x \sim N^{-0.5}$. Correlations between otherwise independent parameters  also induce level repulsion, but Poisson statistics again return at $ \mathcal{O}(N^{-0.5})$ deviations from such correlations. In both cases the crossover occurs roughly over the range $N^{-1.5}\lesssim \delta \lesssim N^{-0.5}$.

Some parameter choices produce matrices that correspond to sectors of certain known quantum integrable models, although general parameter choices do not map to known models. Most important is that, in addition to the linearity in $x$ condition, the ensembles of matrices studied in this work are only constrained by symmetry requirements just like the Gaussian random matrix ensembles. The only difference here is that in the integrable case there are many more symmetries, and they are parameter dependent. We therefore expect our results to apply generally to quantum integrable models with coupling parameters.

Although we justified the numerical results to a certain degree using perturbation theory, an analytic justification for Poisson statistics for integrable matrices is still lacking. Given the relatively simple construction of integrable matrices through basis-independent relations (i.e., matrix equations) involving familiar RMT quantities such as GOE matrices and random vectors\cite{ScYu}, we surmise that an analytic demonstration of our numerical results might  be feasible -- especially in the type-1 case, see, e.g., the discussion below \eref{BCS} and Refs.~\onlinecite{bogo,aleiner}.

 It is interesting to note that  many-body localized\cite{BAA} (MBL) systems are also expected to display Poisson level statistics\cite{OH,PH}, and there exist random matrix ensembles which model localization and its statistical signatures\cite{Shukla,Kr}. Such ensembles are basis dependent, which is natural because localization is a basis-dependent property. The commutation requirements of integrable systems, however, are basis independent, and therefore so is the accompanying integrable matrix theory. A priori, many-body localization and integrability are two independent concepts\cite{mbl}. Despite this fact, integrable matrices do exhibit a parameter-dependent localization property\cite{disorder} in which almost all eigenstates of the matrix $H(x)=xT+V$ are localized in the basis of $V$ for all values of $x$. The stability of this property when a random matrix perturbation is added to $H(x)$, including the possibility of a multifractal phase accompanying the localized and delocalized regimes\cite{Kr}, is the subject of future study.

\begin{acknowledgments}

This work was supported in part by the David and Lucille Packard Foundation and by the National Science Foundation under Grant No. NSF PHY11-25915. E. A. Y. also thanks the University of California at Santa Cruz and Kavli Institute for Theoretical Physics, where a significant part of this research was conducted, for  hospitality. The work at UCSC was supported by the U.S. Department of Energy (DOE), Office of Science, Basic Energy Sciences (BES) under Award \# FG02-06ER46319. We acknowledge Daniel Hansen's  contribution at an early stage of this work, especially to the unfolding technique in Appendix~\ref{appA}. We also thank Joel Lebowitz for helpful discussions.  Finally, we thank the PRE referees for suggestions and questions that led to a considerably improved draft of this work.

\end{acknowledgments}

\appendix

\section{Unfolding spectra}
\label{appA}

The eigenvalue spacing distributions $P(s)$, $\mathcal{P}(s)$ and $p(s)$ (see Sect.~\ref{sec:ergodicity} for definitions) considered in the level statistics data in this work characterize the fluctuations of spacings about their local means. Unfortunately, a non-constant density of states renders the actual spacings inadequate for measuring these fluctuations. Unfolding the spectrum of a matrix refers to the replacement of the actual eigenvalues $\eta_j$ with a new set of numbers with unit mean spacing, but that preserve the character of local fluctuations. 

We employ a simple method that essentially approximates the inverse density of states (i.e., mean level spacing) of a given matrix through linear interpolations. First, we write the eigenvalues $\eta_j$ in increasing order, and express the $j$-th eigenvalue $\eta_j$ in terms of the actual spacings $S_k$
\begin{equation}
\eta_j = \eta_1+ \sum_{k = 0}^{j - 1}{S_k}.
\end{equation}
No unfolding has taken place as of yet,  i.e., this is an exact expression. Now we postulate that we can write the $k^{\mathrm{th}}$ spacing $S_k$ as the product of a smoothly varying local mean spacing $s_k$ and an $ \mathcal{O}(1)$ fluctuating number $\rho_k = 1 + \delta_k$
\begin{equation}
\label{eq:EVs&LevelSpacings}
\begin{split}
\eta_j &= \eta_1+ \sum_{k = 0}^{j - 1}s_k(1 + \delta_k) \\
&= \eta_1+ \sum_{k = 0}^{j - 1}s_k\rho_k.
\end{split}
\end{equation}
Note that the $\rho_k$ have the form of fluctuating numbers with unit mean; they will therefore serve as our unfolded spacings. By their definition we can write them as
\begin{equation}
\rho_k = \dfrac{\eta_{k+1} - \eta_k}{s_k}.
\end{equation}
Therefore, if we calculate the smoothly varying mean level spacings $s_k$ from the given data, we can easily find the unfolded spacings. The ambiguity in our particular unfolding procedure lies in the calculation of $s_k$ because its definition involves choosing how many spacings over which to average, a quantity we call $2r$

\begin{equation}
\label{eq:MeanSpacings}
s_k = \dfrac{\eta_{k+r} - \eta_{k - r}}{2r}.
\end{equation}
It is important to realize that $s_k$ is just the inverse of the density of states. The parameter $r$ is arbitrary except that it must satisfy two conditions:

1) $r$ must be large enough to contain many eigenvalues, which is necessary in order for $s_k$ to be a smooth function of $k$.

2) $r$ cannot be too large or else $s_k$ will actually smooth over features in the true inverse density of states.

In practice we have chosen $r$ to be the floor function of the square root of the maximum number of eigenvalue spacings $\nu$ taken from each matrix. To avoid edge effects we have selected $\nu = 0.8N$. Then $r = \lf\sqrt{0.8N}\rf$. Here are some typical values of $r$ used in this paper

\begin{equation}
\begin{split}
N = 500, & \quad r = 22, \\
N = 1000,  &\quad  r = 31, \\
N = 2000,  &\quad r = 44. \\
\end{split}
\end{equation}
Such a choice of $r$ grows with $N$ but also is small compared with $N$. In other words, we satisfy the requirement $1 \ll r \ll N$ as $N \rightarrow \infty$.

For even the best choices of $r$, our unfolding method can still fail if the inverse density of states varies too quickly or has singularities. Such a situation arises for example in small linear combinations of basis matrices (defined in \eref{eq:Type1Basis} and \eref{genedfin}) for any type. Consider, for example, the insets of Figs.~\ref{1matrixN20000M1basis} and \ref{ansatz basis} that show the densities of state of integrable basis matrices. The large portions of the spectra where the density of states $D(\eta)$ drops to zero is typical of small linear combinations of basis matrices. This behavior is generic for basis matrices of all types, and it can be understood by first considering the expression for the eigenvalues of a type-1 basis matrix (in the primary parametrization) where $d_k = \delta_{k,q}$
\begin{equation}
\label{eq:BasisType1EV}
\eta_j = \dfrac{\gamma_q^2}{\lambda_j - \varepsilon_q}.
\end{equation}
As both the $\lambda_j$ and $\varepsilon_j$ have finite support, $\eta_j$ in this case will only approach within a finite distance of zero. 

An analogous argument exists for basis matrices in the ansatz parametrization for any type. For linear combinations of a small number of basis matrices, such gaps may overlap, but a mean level spacing $s_k$ will still be ill-defined in many parts of the spectrum. As the number of basis matrices in the linear combination increases, the gaps smooth out until $s_k$ is well-defined everywhere. 

Given such gaps in spectra, no choice of $r$ will give the consistent level statistics. This can be seen numerically by varying $r$ and observing that $P(s)$ is strongly dependent on $r$. We must then avoid regions of the spectrum where $1/D(\eta)$ is poorly behaved.

The difficulty in this task is to automate it so that we can unfold many matrices in succession without having to examine each one by hand. Our solution is to notice that if there are a small number of spacings in the spectrum that are many times the local mean spacing, the standard deviation of the set of actual spacings will be large. If the standard deviation (normalized by the mean) of the actual spacings is near unity, we can be sure that there are no huge jumps such as the ones in Figs.~\ref{1matrixN20000M1basis} and \ref{ansatz basis}.

With these considerations, here is our unfolding algorithm:

(1) Calculate $SD = \dfrac{\textrm{Standard Deviation}}{\textrm{Mean}}$ of the middle $(80\% + 2r)$ of the spectrum's actual spacings. If $SD < 1.2$, unfold this region of the spectrum with $r = \lf\sqrt{0.8N}\rf$ and continue to next matrix. If not, continue to step (2).

(2) Shift the region of the spectrum in question to the right by 10 eigenvalues.

(3) If $\eta_{\textrm{max}} > \eta_{0.9N}$, use 10 fewer spacings AND restart $\eta_{\textrm{min}} = \eta_{0.1N}$

(4) Calculate SD. If $SD > 1.2$, repeat back to step (2). If $SD < 1.2$ unfold this region of the spectrum with $r = \lf\sqrt{0.8N}\rf$ and continue to next matrix.

This procedure allows for fewer than $0.8N$ of the spacings to be used, but we are guaranteed to only investigate regions of the spectrum where the mean level spacing accurately represents the size of a typical spacing. Once a sufficiently large number of basis matrices are used in linear combination, the entire middle $80 \%$ of the spectrum behaves smoothly and the procedure given above terminates at step (1) for each matrix. The choice of a maximum SD of 1.2 is somewhat arbitrary, and in some parts of this work we used 1.5 in order to increase computation speed (i.e., keep more eigenvalue spacings per matrix). Apart from slight differences in distributions, our results are unaffected by this arbitrariness.

The unfolding procedure used in this paper assumes that the level statistics are the same in all regions of the spectrum, excluding the edges. Although in principle a Hermitian matrix can have different spectral statistics in different parts of its spectrum,  we numerically showed in Sect.~\ref{sec:ergodicity} that the statistics are the same in all parts of the spectrum of integrable matrices $H(u)$, i.e., they are translationally invariant.


\begin{thebibliography}{99}
\bibitem{note2}  In this Introduction, by quantum integrable systems, models, or Hamiltonians we mean exactly solvable many-body models  also colloquially   called integrable. On the other hand, an `integrable matrix' is a precise concept  defined in the 4-th and 5-th paragraphs.
\bibitem{note3} One can equivalently attach the parameter to the $V$-part, i.e., write $H(u)=T+uV$ as in some of the previous work, or introduce two parameters, $H(x,u)=xT+uV$, see Ref.~\onlinecite{ScYu} for more detail.
\bibitem{note1} Two matrices are considered equal if they differ by a multiple of the identity $(c_1x+c_2)\iden$, which is not included in $H^i(x)$. For example, linear independence means that $\sum_i a_i H^i(x)=(c_1x+c_2)\iden$ iff $a_i=0$.
\bibitem{poissonnote} Any ensemble composed of lists of independent random numbers is referred to as the ``Poisson'' ensemble.
\bibitem{rigiditynote} Berry works with the spectral ridigity, which is a statistic closely related to the number variance.
\bibitem{asterisk} In this work, the $g_j$ indeed are eigenvalues from an $N\times N$ GOE matrix, but are selected as a set of $N-M$ consecutive eigenvalues instead of $N-M$ randomly chosen ones. This choice does not affect the level statistics.
\bibitem{mbl} MBL  seems to be a broader class. Even though certain integrals of motion have been identified in MBL systems (see Ref.~\onlinecite{SPA} for example), they have not been shown to be exactly solvable or integrable as defined in this paper.
\end{thebibliography}
\end{document}